\documentclass[draft]{agujournal2019}

\usepackage{cleveref}
\Crefname{figure}{Fig.}{Figs.}
\usepackage{graphicx}
\usepackage{tabularx}
\usepackage{siunitx}

\usepackage{url} 
\usepackage{lineno}
\usepackage[inline]{trackchanges} 
\usepackage{soul}
\usepackage{listings}

\usepackage{subcaption}
\usepackage{svg}

\addeditor{FB}
\addeditor{FL}
\addeditor{CD}
\addeditor{ClDy}

\journalname{Space Weather}

\begin{document}

%
%


\title{AniMAIRE - A New Openly Available Tool for Calculating Atmospheric Ionising Radiation Dose Rates and Single Event Effects During Anisotropic Conditions.}

%
%




\authors{C. S. W. Davis\affil{1}, F. Baird\affil{1}, F. Lei\affil{1}, K. Ryden\affil{1},C. Dyer\affil{1,2}}


\affiliation{1}{Surrey Space Centre, University of Surrey, Guildford, U.K.}
\affiliation{2}{CSDRadConsultancy}




\correspondingauthor{C. S. W. Davis}{ChrisSWDavis@gmail.com}




\begin{keypoints}
\item AniMAIRE can simulate atmospheric radiation and electronics effects from any inputted anisotropic distribution of particles hitting Earth. 
\item AniMAIRE predicts GLE71 dose rates between those of CRAC:DOMO and WASAVIES, and event-integrated dose rates are found to require anisotropy.
\item AniMAIRE has been made openly available online, such that anyone can run and experiment with it, or contribute to the model through forking. 
\end{keypoints}

%
%

%
%


\begin{abstract}
AniMAIRE (Anisotropic Model for Atmospheric Ionising Radiation Effects) is a new model and Python toolkit for calculating radiation dose rates experienced by aircraft during anisotropic solar energetic particle events. AniMAIRE expands the physics of the MAIRE+ model such that dose rate calculations can be performed for anisotropic solar energetic particle conditions by supplying a proton or alpha particle rigidity spectrum, a pitch angle distribution, and the conditions of Earth's magnetosphere. In this paper, we describe the algorithm and top-level structure of AniMAIRE and showcase AniMAIRE's capabilities by analysing the dose rate maps that AniMAIRE produces when the time-dependent spectra and pitch angle distribution for GLE71 are inputted. We find that the dose rates AniMAIRE produces for the event fall between the dose rates produced by the WASAVIES and CRAC:DOMO models. Dose rate maps that evolve throughout the event are also shown, and it is found that each peak in the inputted pitch angle distribution generates a dose rate hotspot in each of the polar regions. AniMAIRE has been made available openly online so that it can be downloaded and run freely on local machines and so that the space weather community can easily contribute to it using Github forking. 
\end{abstract}

\section*{Plain Language Summary}
Occasionally an eruption occurs on the surface of the Sun, ejecting high energy particles into the solar system in what are known as solar particle events. If this eruption of particles happens to hit Earth and have a sufficiently high kinetic energy, the particles can penetrate Earth's magnetosphere and atmosphere, creating particle radiation showers that increase radiation dose rates in Earth's atmosphere. These situations are known as Ground-Level Enhancements (GLEs), which occur once every few years on average. GLEs can cause issues in aircraft electronics and cause passengers and crew on airplanes to experience heightened radiation doses. In this paper, we describe a new, openly available model we've created to calculate dose rates in the atmosphere during GLEs, called AniMAIRE. Most models that exist to date are only able to calculate radiation dose rates when all particles are hitting Earth equally from all directions, however this approximation is generally only reliable during the later stages of events, and isn't accurate for many GLEs. AniMAIRE has therefore been designed so that it doesn't need to rely on this approximation, and can calculate dose rates for situations where it is supplied a direction-dependent particle flux. In this paper we also show and analyse the dose rate maps AniMAIRE produces when running across a GLE in May 2012, finding several interesting results.

%
%

%


%
%
%
%

\section{Introduction}

Atmospheric radiation dose rates and Single Event Effects (SEEs) induced by Ground Level Enhancements (GLEs) are one of the numerous hazards that space weather can create on Earth \cite{shea2012space,meier2020radiation,dyer1989measurements,dyer1990measurements,tobiska2015advances}. GLEs occur on average once every few years \cite{karapetyan2008detection, usoskin2015database} and can cause large increases in radiation dose rates versus normal `quiet' cosmic-ray-only conditions  \cite{dyer2007solar,copeland2008cosmic}; for example, increases by factors of 1000 or more versus quiet conditions could be expected on a time scale of between 1 in 40 and 1 in 70 years  \cite{dyer2017extreme}. The UK Royal Academy of Engineering predicted that a large GLE could cause airline crew and passengers to receive dose rates that exceed 20 times the recommended general public exposure limit  \cite{cannon2013extreme}. Radiation from these events could also cause electronics issues in aircraft systems due to Single Event Effects (SEEs)   \cite{normand1993altitude, taber1993single}, potentially increasing pilot workloads if issues were to occur in multiple systems at once. It is self-evident that lowering the risk of these issues occurring is important to both the airline industry as a whole and airline passengers. Therefore, simulating the dose rates and SEEs an aircraft will experience during a large space weather event is an important part of both understanding the dose rates aircraft will experience, and developing dose rate mitigation strategies.

The simulation of atmospheric radiation levels is currently an active area of research, and many models have been developed over the last few decades to model atmospheric radiation. Many models have historically focused on simulating cosmic ray quiet time dose rates. However, recent modeling work in the field has moved towards simulating GLEs, although typically with assumptions of isotropy. In this paper, we discuss the development and release of a new and openly available Python model for simulating anisotropic dose rates due to incoming protons - called AniMAIRE. AniMAIRE is based on the physics behind its precursor model, MAIRE+ \cite{hands2022new}, but with the additional feature that it can simulate dose rates during anisotropic conditions. 

The MAIRE+ model \cite{hands2022new} was developed by the Space Environment and Protection Group at Surrey Space Centre in collaboration with the UK Met Office as part of the UK's SWIMMR programme. MAIRE+ is a successor model to the previous QARM \cite{lei2004atmospheric,lei2005improvements}, and MAIRE \cite{MAIREUK,hands2016new} models, using many of the same particle responses that were present in these. It is an isotropic model with input conditions tailored for the UK airspace. Other major models that can currently perform calculations during isotropic conditions include CARI-7  \cite{copeland2017cari}, AVIDOS  \cite{latocha2009avidos} and NAIRAS  \cite{mertens2009development,mertens2013nairas}. 

While isotropic models are more accurate during the later stages of most GLEs, they are unable to accurately model dose rates during more anisotropic stages of GLEs (typically the initial 30 minutes to 60 minutes of a GLE, although the stages and features of a GLE vary considerably from event to event \cite{shea2012space}). The only currently used major models that we are aware of that can determine dose rates during anisotropic conditions are WASAVIES  \cite{kataoka2014radiation, sato2018real}, SIGLE \cite{lantos2003history,lantos2004semi} and CRAC:DOMO by \citeA{mishev2015computation}. SIGLE takes an empirical approach to dealing with anisotropy, using the count rate increases at the nearest ground-level neutron monitor to a particular flight to apply dose rate weighting for anisotropy.

In contrast, WASAVIES takes a physics-based approach and effectively traces a full simulation of solar particle injection profiles through the solar heliosphere to Earth's magnetosphere. WASAVIES then traces particle directions through Earth's magnetosphere using the Tsyganenko 1989 model \cite{tsyganenko1989magnetospheric}, and then converts the particle spectra hitting Earth's atmosphere into dose rates. It does all this using a combination of pre-calculated response databases and real-time simulations. Fitting against experimental measurements is used to determine the most appropriate injection profile for a particular event.

The CRAC:DOMO model is similar to WASAVIES and also uses trajectory tracing through Earth's magnetosphere using the Tsyganenko 1989 model (implemented through PLANETOCOSMICS  \cite{PLANETOCOSMICS,desorgher2006planetocosmics}) to convert particle distributions at Earth's magnetosphere into dose rates. The key difference between CRAC:DOMO and WASAVIES here is that CRAC:DOMO assumes nothing about the heliosphere and instead spectra and pitch angle distributions at Earth's magnetosphere are directly inserted into the model, which are determined through a separate fitting model using ground-level neutron monitor data.


Few atmospheric radiation models (isotropic or anisotropic) have been released openly both in terms of running and for modification in general, which is likely one of the reasons why so many models have been developed in general, as every group in the field has so far had to create their own model to perform research. One positive recent development has been the recent addition of models to NASA CCMC \cite{CCMC} system, for instance, NAIRAS \cite{mertens2023nairas}, for real-time running, and the development of ESA's Network of Models (NoM) \cite{NoM} where models can be run on-demand.

To aid the atmospheric radiation community, AniMAIRE has currently been released openly \cite{AniMAIRE} on Github and PyPi, where it can be installed with a single command using the most common Python package installer, `pip' and run locally by any users with a moderate level of experience in Python. Links to AniMAIRE and all of the software it depends on (including several additional useful software packages that were developed as part of this work) can be found in the Software and Data Availability section at the end of this paper. One of the reasons why Python has become so popular in the last decade in the scientific community as a whole is because of the ease at which outputs from Python `modules' such as AniMAIRE developed by users can be fed into each other, and as AniMAIRE has been made open source it can be used in other projects in the community as a dependency.

The source code of AniMAIRE is openly available for users who wish to investigate the algorithm in more detail or for members of the community who wish to make their own contributions and improvements to the software through Github `forking' and pull requests. We are also currently investigating adding AniMAIRE to the CCMC \cite{CCMC} and NoM \cite{NoM} so that users without Python experience can also perform calculations with AniMAIRE. More details on how to access and install AniMAIRE and all its dependencies can be found in the data availability section at the end of this paper.

AniMAIRE therefore provides novel features to the community of being openly accessible, able to perform anisotropic calculations, a modular software design, and the ability to be run locally and as part of other future software. 

\section{The AniMAIRE Physics/Algorithm}
\label{sec:physics}
Like MAIRE+, AniMAIRE uses altitude-dependent yield functions to convert an energetic particle spectrum to a radiation dose rate at a given location on Earth. However, when the SEP spectrum is anisotropic, the spectrum which can be `seen' from a given location on Earth is specific to that location. Therefore, AniMAIRE calculates the energetic particle spectrum, $J(P)$ for a given location on Earth as
\begin{equation}
J(P) = F(P) \times J_\parallel(P) \times G(\alpha(P))
\label{eq:ep_spectrum}
\end{equation}
$J_\parallel(P)$ is the SEP proton spectrum arriving along the `axis of symmetry', \textit{i.e.} the spectrum at the direction of maximum flux, as a function of particle rigidity $P$, which is usually close to the Interplanetary Magnetic Field (IMF) direction \cite{mishev2016analysis,mishev2021application}. The anisotropy is accounted for through the pitch angle distribution, $G(\alpha(P))$.  The parameter $\alpha(P)$ is the pitch angle between the asymptotic directions corresponding to the given location and the location of the axis of symmetry. Since the asymptotic direction at a given location is a function of the rigidity of the particle reaching that location, the pitch angle is also a function of particle rigidity. Finally $F(P)$ is a filtering function, with a value of 0 for forbidden values of $P$ and 1 for allowed values (see \citeA{Cooke1991} for an overview of asymptotic directions and cut-off values). 

$J_\parallel(P)$ can also be set to be the cosmic ray spectrum, which can be calculated in AniMAIRE through the DLR-modified ISO model  \cite{matthia2013ready} as implemented in the CosRayModifiedISO Python module  \cite{cosraymodifiediso}. In this cosmic ray-only case, $G(\alpha(P))$ is set to be isotropic, i.e. always equal to 1.

The general AniMAIRE process for calculating dose rates and fluxes at a single location is as follows:

\begin{enumerate}
    \item Asymptotic directions are calculated for rigidities between 0.1~GV and 1010.0~GV using MAGNETOCOSMICS \cite{MAGNETOCOSMICS,desorgher2004magnetocosmics} through the AsympDirsCalculator Python module \cite{AsympDirsCalculator}. Currently, only the Tsyganenko 1989 model \cite{tsyganenko1989magnetospheric} can be used, but more models could be added in future versions of AsympDirsCalculator.
    \item The pitch angle between each asymptotic direction and the axis of symmetry is calculated using the dot product of the asymptotic direction with the incoming direction of the solar particle event. The pitch angle  can be calculated using the equation $\cos(\alpha) = \sin(\Lambda)\sin(\theta_s) + \cos(\Lambda) \cos(\theta_s) \cos(\Phi-\phi_s)$ given by \citeA{cramp1997october}. The pitch angle distribution is assumed to be symmetric about this particular direction. 
    \item The full energetic particle spectrum is calculated using \cref{eq:ep_spectrum}.
    \item The full energetic particle spectrum is multiplied by the relevant yield functions for each of the output quantities and integrated with respect to rigidity. This is achieved through the atmosphericRadiationDoseAndFlux Python package \cite{atmosphericRadiationDoseAndFlux}.
\end{enumerate}

\begin{figure}
    \centering
    \includegraphics[width=\textwidth]{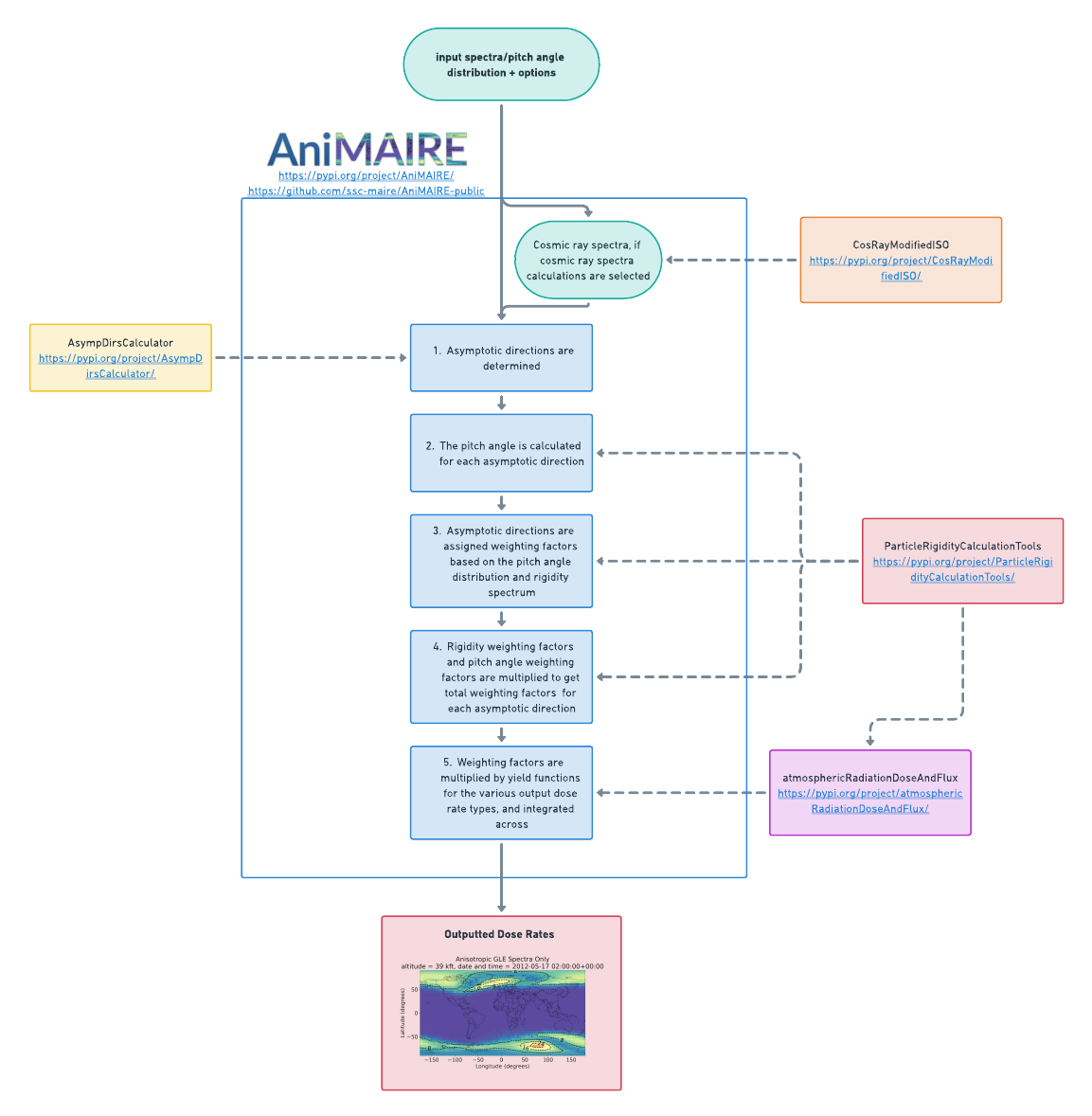}
    \caption{A diagram showing the general algorithm AniMAIRE uses, as well as the dependency packages that were created as part of the AniMAIRE development process.}
    \label{fig:AniMAIREdiagram}
\end{figure}

A diagram of this procedure is shown in \cref{fig:AniMAIREdiagram}. 

A standard run of AniMAIRE takes as input any generic proton and/or alpha particle spectrum, a list of latitude/longitude pairs, a list of altitudes, pitch angle distributions, Kp index, the date and time for the run to correspond to, and reference pitch angle coordinates (the location on Earth where the pitch angle is 0.0, i.e. the direction in which the GLE is hitting Earth from). Many of these input parameters have default values so that a user doesn't have to specify every parameter all the time - for instance, if only a proton spectrum is specified, AniMAIRE will automatically calculate the Kp index using the PySpaceWeather Python module \cite{PySpaceWeather} using the specified date and time. Equally, if a date and time is not specified, AniMAIRE will use the current date and time by default.

Spectra do not always have to be specified directly. AniMAIRE can calculate dose rates from incoming cosmic rays as stated earlier, or through supplying parameters for several standard GLE proton spectrum distributions. Examples of using these additional methods can be found at AniMAIRE's Github and PyPi pages, and AniMAIRE could be relatively trivially expanded in the future by members of the space weather community so that it can be directly run using spectra from other models.


\section{Application to Cosmic Rays and GLE71}

Links to all of the data and Jupyter Notebooks used to generate the plots in this section can be found in the Software and Data Availability section at the end of this paper.

To showcase the outputs that AniMAIRE can produce for anisotropic situations, AniMAIRE was run across the time-dependent spectra and pitch angle distributions for GLE71 reported in \citeA{mishev2021application}. GLE71 was a complex event, and Mishev et al. found that it was best modeled using a double Gaussian pitch angle distribution with a modified power law rigidity spectrum. They also found that the resulting distribution agreed well with SEP measurements from the Payload for Antimatter Matter Exploration and Light-nuclei Astrophysics (PAMELA)\cite{bruno2018solar} when integrated across the whole event.

The spectra and pitch angle distribution for each time interval were fed into AniMAIRE to produce dose rate maps at each point in time throughout GLE71. The input rigidity spectra equation corresponding to the GLE used was:

\begin{equation}
    J_\parallel(P) = J_0 P^{-(\gamma +\delta\gamma(P-1) )}
    \label{eq:rigSpectrum}
\end{equation}

and the pitch angle distribution used was:


\begin{equation}
    G(\alpha(P)) =\exp{\bigg(-\frac{\alpha^2}{\sigma_1^2}\bigg)} + B \exp{\bigg(-\frac{(\alpha - \alpha^\prime)^2}{\sigma_2^2}\bigg)}
    \label{eq:pitchAngle}
\end{equation}

Here, $P$ is the rigidity in units of GigaVolts. Note that the pitch angle itself is a function of rigidity $P$ and is defined relative to the axis of symmetry of the incoming SEP event, given with $\Phi$ and $\Lambda$. These parameters, as well as $\gamma$, $\delta\gamma$, $\sigma_1$, $\sigma_2$ and $\alpha^\prime$ were given by Mishev et al. and were determined through their fitting method - although $\gamma$ values in the table are listed as negative and have to be multiplied by $-1$ to insert into \cref{eq:rigSpectrum}.

\begin{figure}
    \centering
    \includegraphics[width=\textwidth]{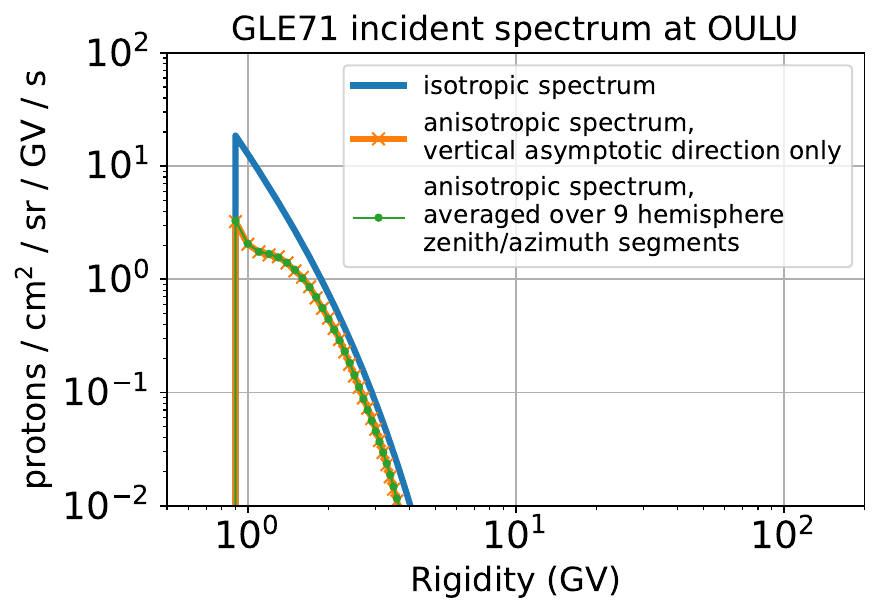}
    \caption{The GLE spectra for several different cases that are incident upon a latitude of 65.02\textdegree~ and a longitude of 25.5\textdegree, the location of the OULU neutron monitor at the first timestamp for the event. The vertical anisotropic spectrum is a direct multiplication of pitch angle values with the isotropic spectra, which is why the anisotropic spectra are always less than the isotropic spectrum here (the pitch angle distribution used here was always less than 1, with a very conservative isotropic spectrum).}
    \label{fig:spectraPlotted}
\end{figure}

An example of the effect of multiplying the anisotropic spectra and pitch angle distribution together is shown in \cref{fig:spectraPlotted}, which shows the spectra for three different cases which impact the latitude of 65.02\textdegree~and longitude of 25.5\textdegree~in the first timestamp of GLE71. The isotropic spectrum is the raw spectrum, $J_\parallel(P)$, from \cref{eq:rigSpectrum}, and the anisotropic spectra are the spectra multiplied by the pitch angle distribution, as calculated by AniMAIRE, i.e. $J(P)$ from \cref{eq:ep_spectrum}. The latitude and longitude used were chosen somewhat arbitrarily for this comparison but are the coordinate of the commonly used OULU neutron monitor.

The anisotropic spectra are less than the isotropic spectrum at all points as the value of the pitch angle distribution ($G(\alpha)$) is always less than approximately 1 in this case. The anisotropic spectra are also different in shape to the isotropic spectrum at lower rigidities, illustrating the complex effect of the pitch angle distribution on the spectra a coordinate receives.

\begin{figure}
    \centering
    \begin{subfigure}[b]{0.85\linewidth}
        \includegraphics[width=\textwidth,trim={0 3cm 0 4cm},clip]{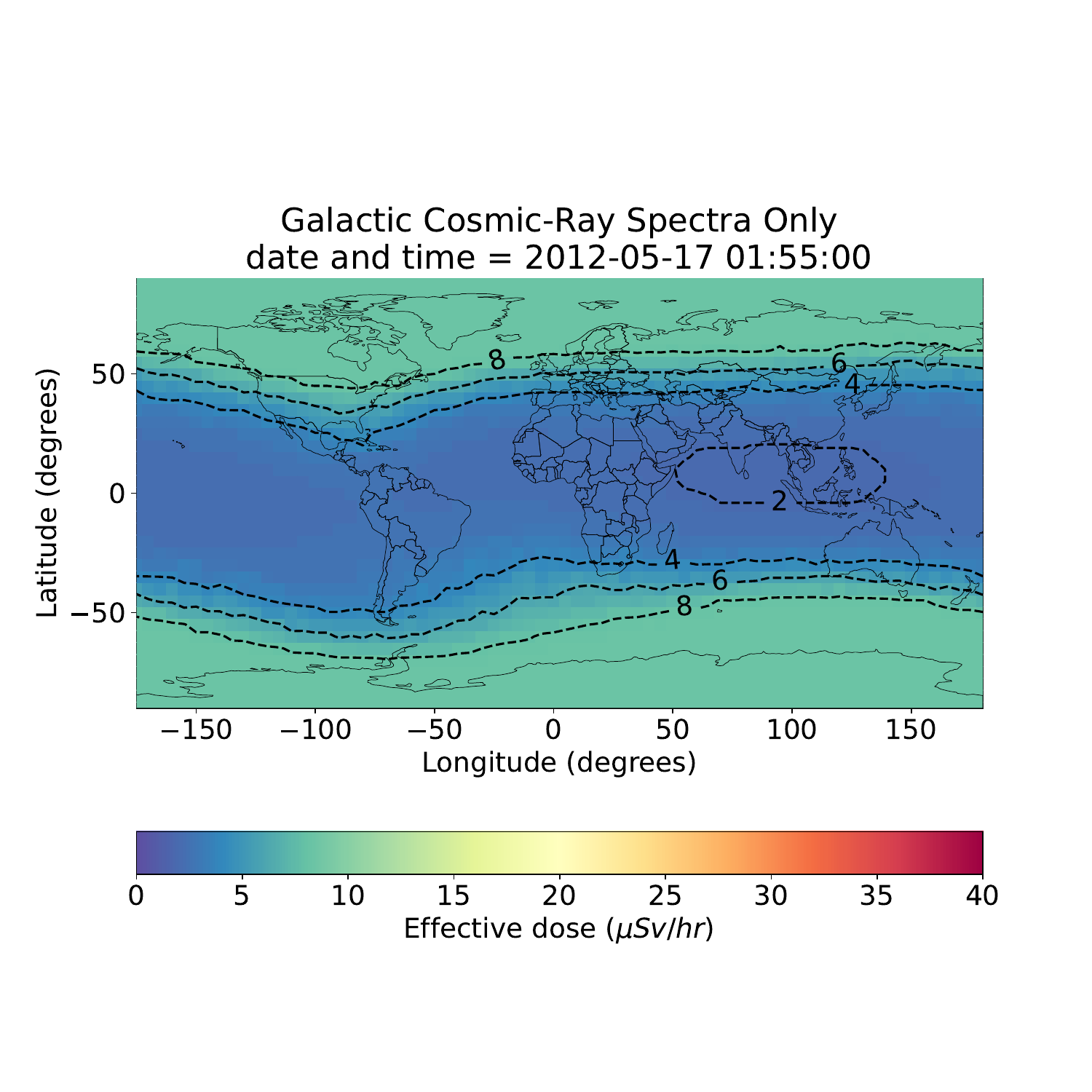}
        \caption{}
        \label{fig:CosRayMap}
    \end{subfigure}
    \begin{subfigure}[b]{0.85\linewidth}
        \includegraphics[width=\textwidth,trim={0 3cm 0 4cm},clip]{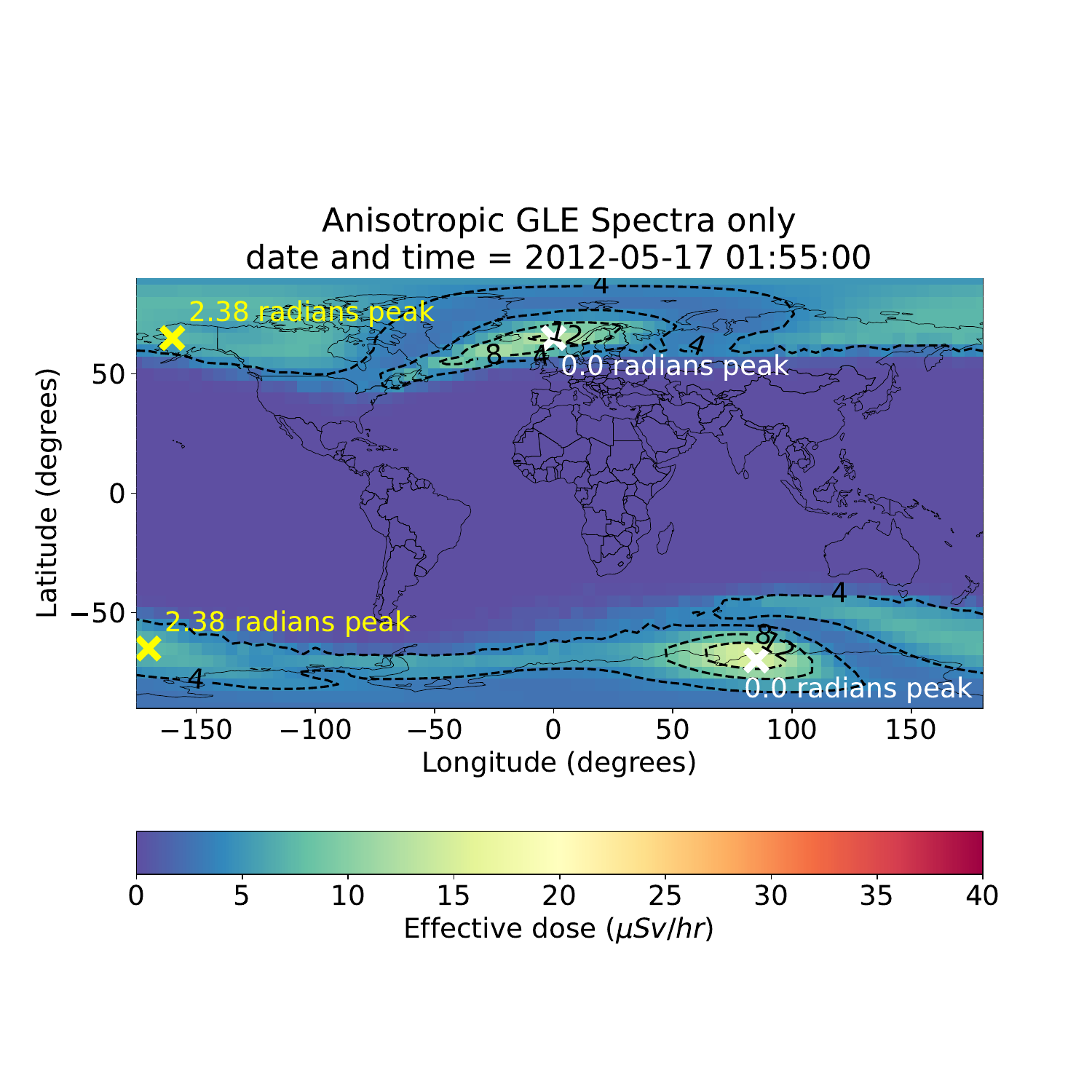}
        \caption{}
        \label{fig:GLE_single_map}
    \end{subfigure}
    \caption{Dose rates at 39~kft in altitude due to galactic cosmic rays are shown in \cref{fig:CosRayMap}. The dose rates follow the structure of Earth's magnetic vertical cut-off rigidities, peaking in the polar regions. Dose rates due to the initial anisotropic GLE spectrum are shown in \cref{fig:GLE_single_map}. The crosses on \cref{fig:GLE_single_map} are the locations where the connection of 1.5~GV protons to the 0.0~radian pitch angle peak (coloured in white) and 2.38~radian pitch angle peak (coloured in yellow) was found to be maximal for the north and south hemisphere.}
    \label{fig:SingleMaps}
\end{figure}

Dose rate maps for the first timestamp in GLE71 were generated for both cosmic rays and anisotropic input spectra and are shown in \cref{fig:SingleMaps}, and the effect of anisotropy on the spatial distribution of dose rates can be seen in \cref{fig:GLE_single_map} versus the isotropic cosmic ray dose rates. Polar regions of heightened dose rates are the same approximate shape and location as in an isotropic situation but contain a complex internal geographic structure. The double Gaussian pitch angle distribution that was used in this event causes two subregions of higher dose rates in the polar region of each hemisphere. The location of the peak dose rate in each of these subregions corresponds to the location with the strongest connection through the asymptotic direction to one of the peaks. The subregions corresponding to the 0.0~radian pitch angle peak contain the most intense dose rates, which is to be expected as the 0.0~radian pitch angle peak represents the strongest pitch angle weighting factors. While the subregions corresponding to the second pitch angle peak at 2.38~radian contain lower dose rates, they encompass a wider area, which is likely because the peak is broader than the first.

An effective dose rate map for the initial phase of GLE71 has been previously been calculated using CRAC:DOMO and presented in Figure~5 of \citeA{mishev2015computation} for 39~kft. This dose rate map shows some similarities to \cref{fig:GLE_single_map}, but also some differences, likely caused by a combination of factors. The southern hemisphere regions show the same 0.0 radian pitch angle peak dose rate feature between 50\textdegree~ and 100\textdegree; however, the map by Mishev et al. shows no obvious sharp dose rate peaks in the northern hemisphere. A key mapping difference here is that Mishev et al. used a smoothed 5\textdegree~ by 15\textdegree~ resolution, in contrast to the 5\textdegree~ by 5\textdegree~ resolution in \cref{fig:GLE_single_map}, so it is likely that many of the granular features that can be seen in \cref{fig:GLE_single_map} have been missed in the map by Mishev et al. Also note that as is discussed later in this paper, the AniMAIRE dose rate yield functions likely differ from those used by CRAC:DOMO.

\begin{figure}
    \centering
    \begin{subfigure}[t]{0.32\linewidth}
          \centering
          \includegraphics[width=\textwidth]{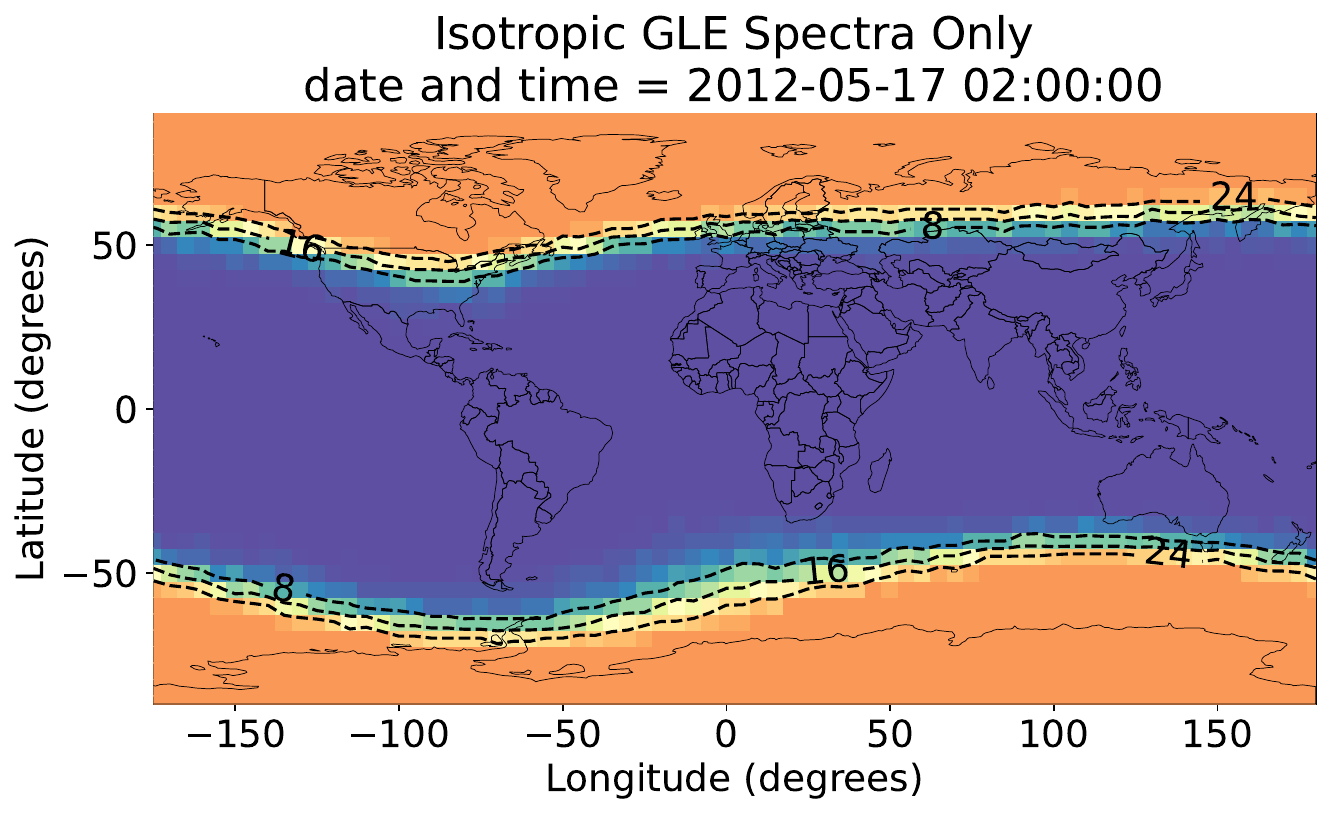}
          \caption{}
          \label{fig:sfig1}
    \end{subfigure}
    \begin{subfigure}[t]{0.32\linewidth}
          \centering
          \includegraphics[width=\textwidth]{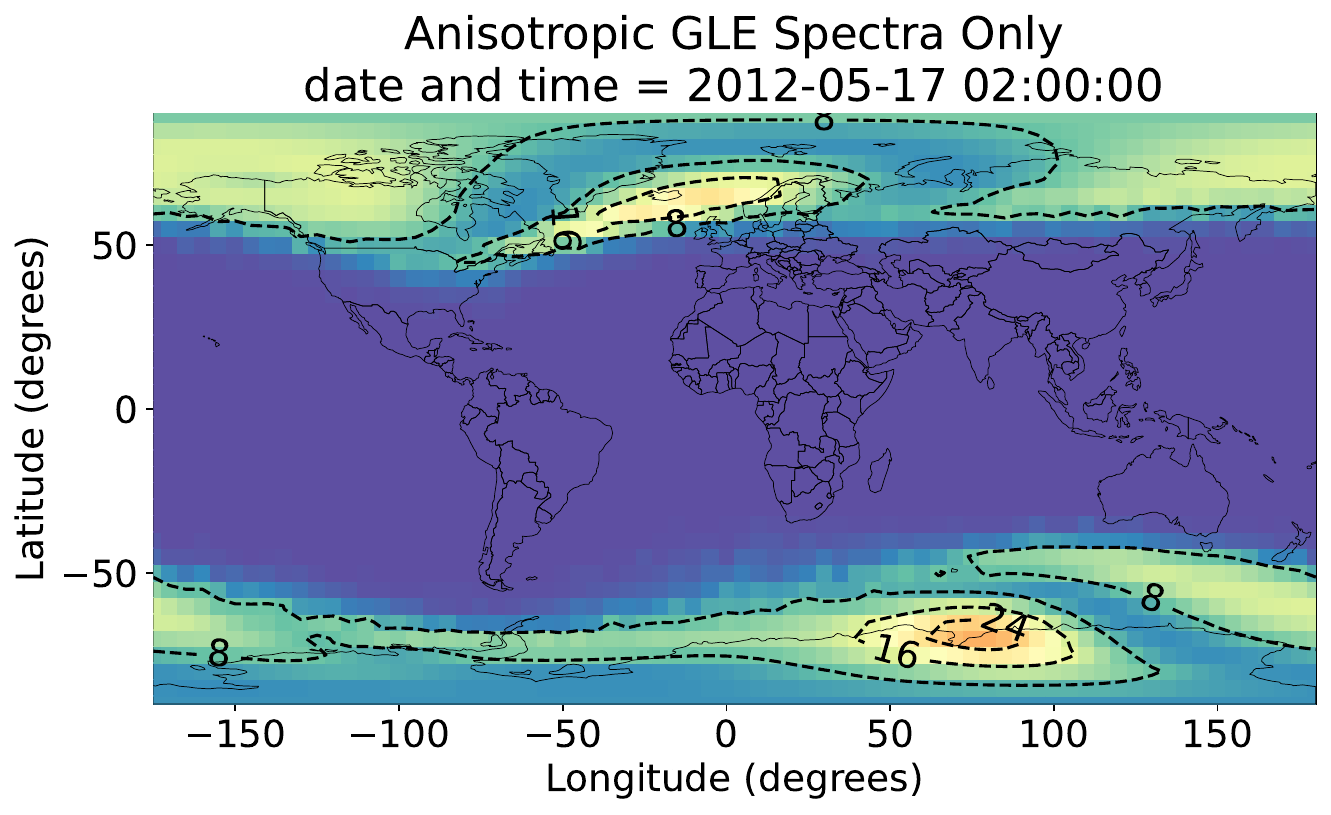}
          \caption{}
          \label{fig:sfig1}
    \end{subfigure}
    \begin{subfigure}[t]{0.32\linewidth}
          \centering
          \includegraphics[width=\textwidth]{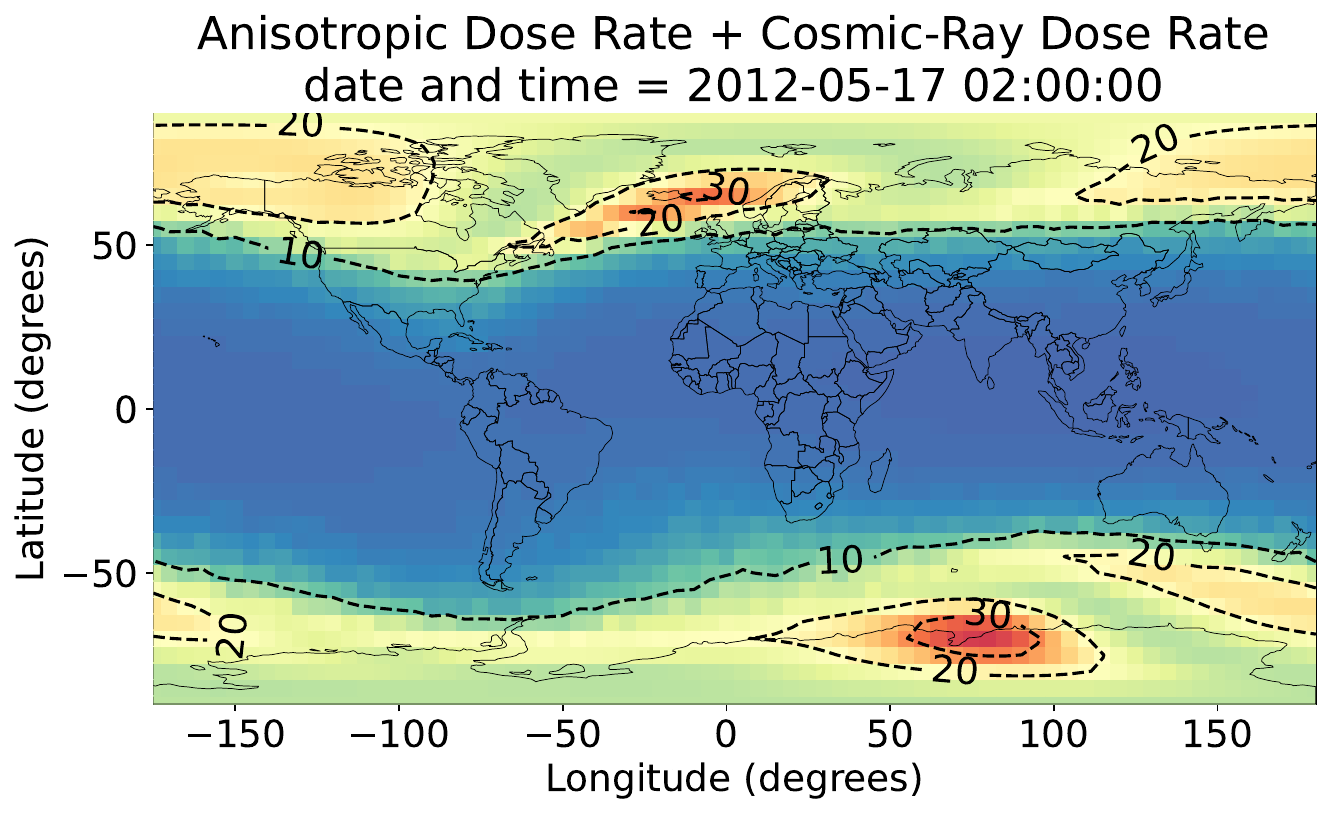}
          \caption{}
          \label{fig:sfig1}
    \end{subfigure}

    \begin{subfigure}[t]{0.32\linewidth}
          \centering
          \includegraphics[width=\textwidth]{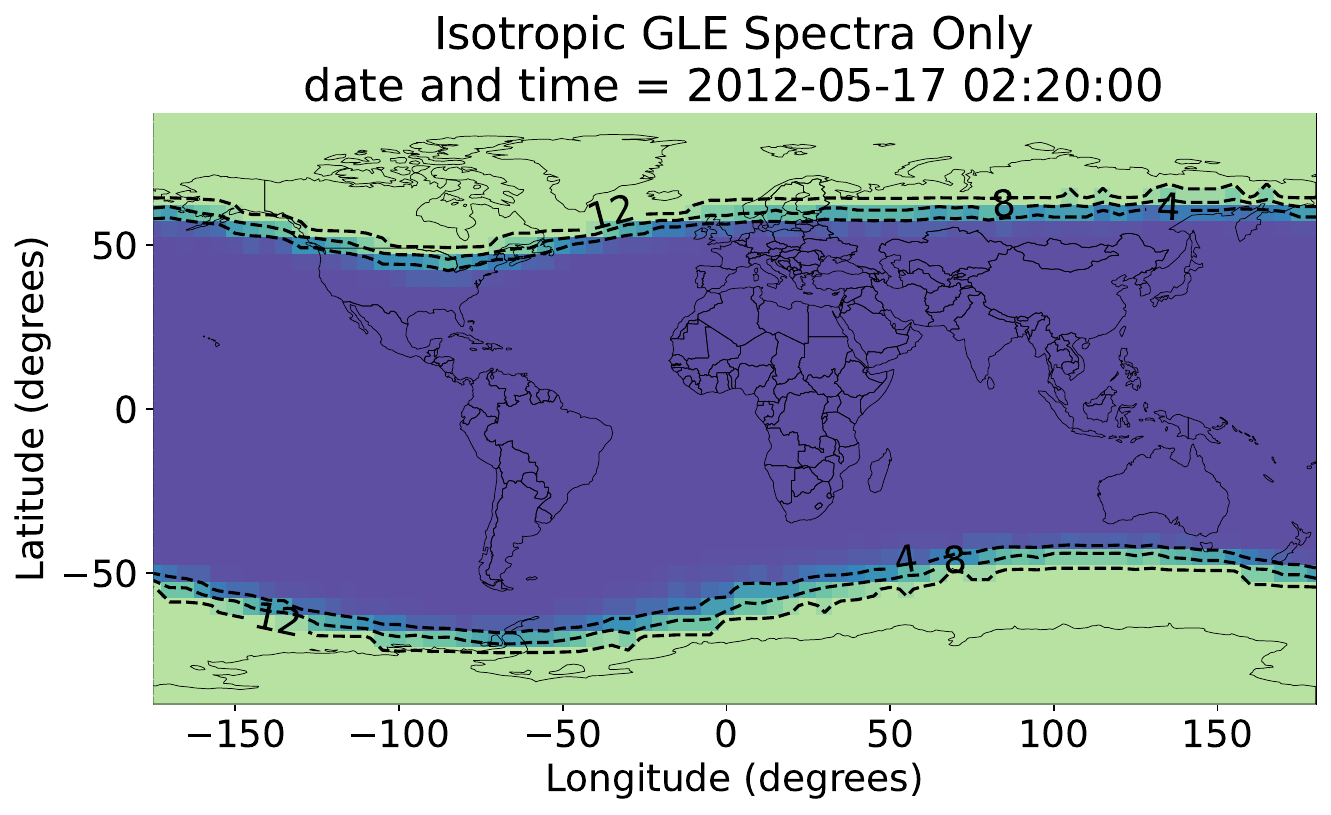}
          \caption{}
          \label{fig:sfig1}
    \end{subfigure}
    \begin{subfigure}[t]{0.32\linewidth}
          \centering
          \includegraphics[width=\textwidth]{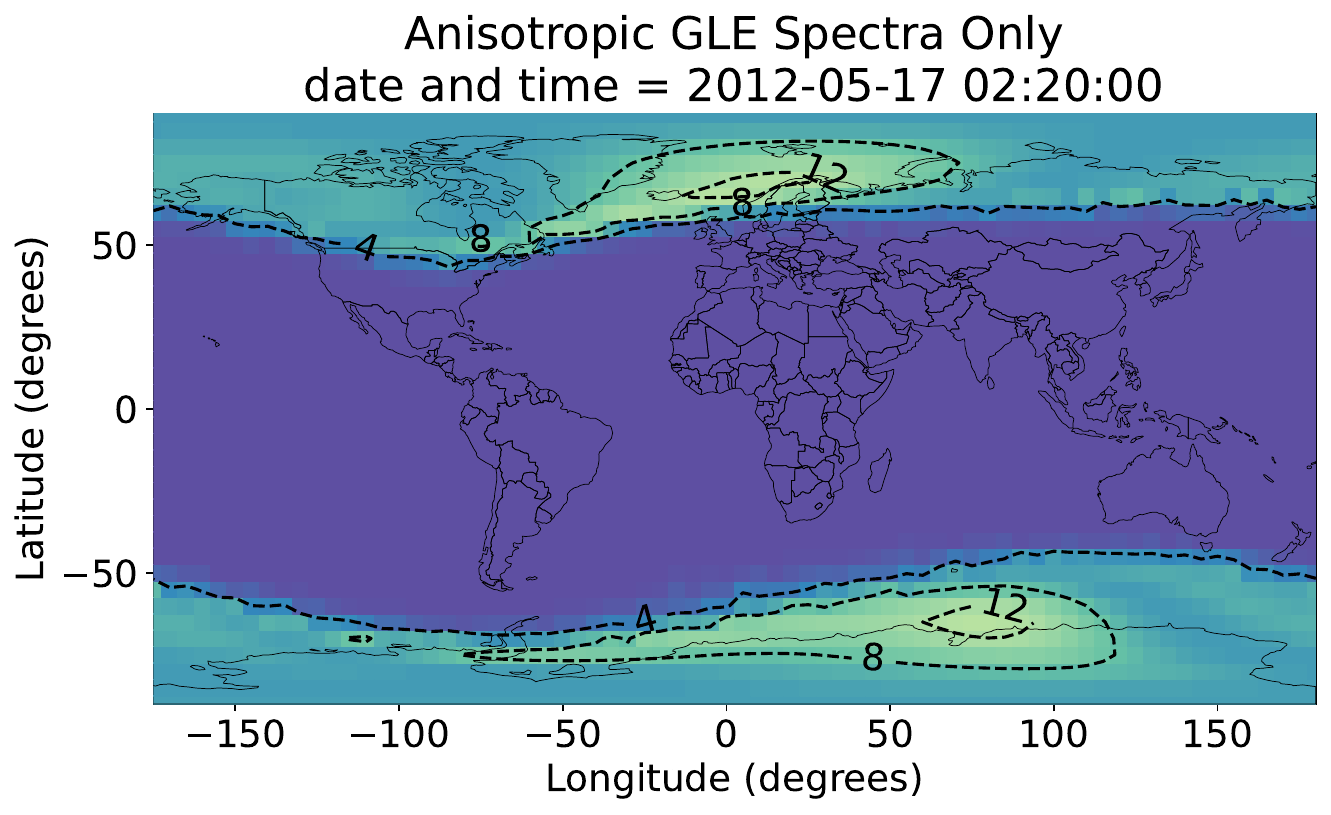}
          \caption{}
          \label{fig:sfig1}
    \end{subfigure}
    \begin{subfigure}[t]{0.32\linewidth}
          \centering
          \includegraphics[width=\textwidth]{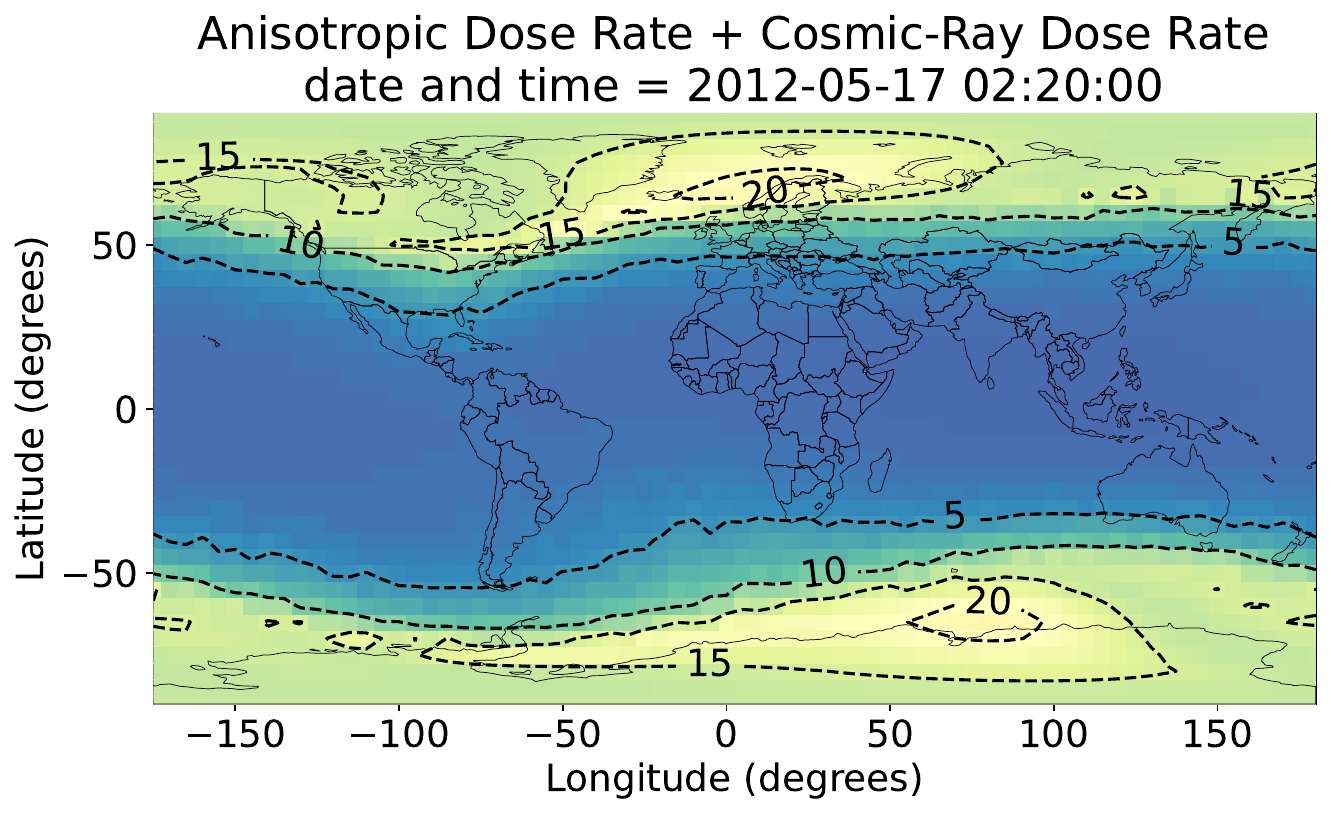}
          \caption{}
          \label{fig:sfig1}
    \end{subfigure}

    \begin{subfigure}[t]{0.32\linewidth}
          \centering
          \includegraphics[width=\textwidth]{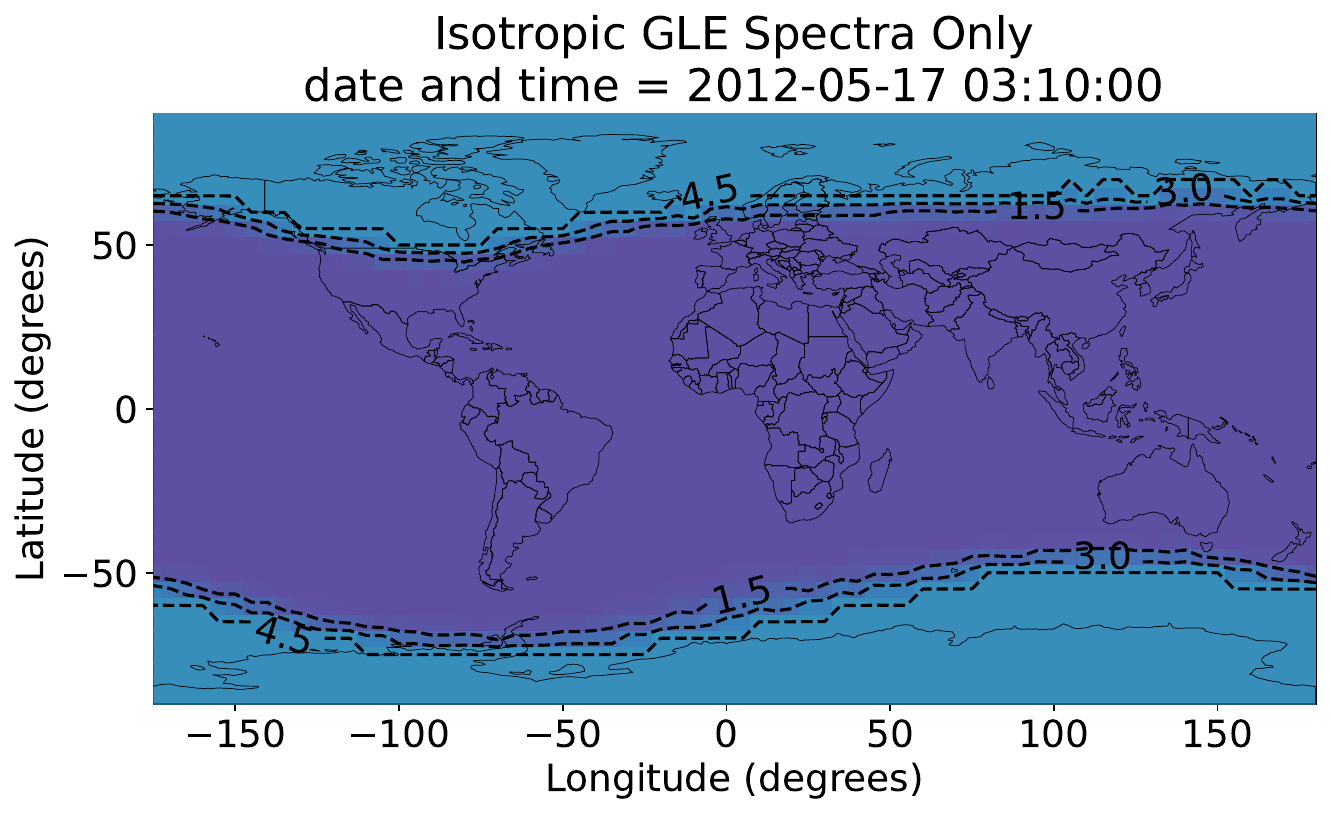}
          \caption{}
          \label{fig:sfig1}
    \end{subfigure}
    \begin{subfigure}[t]{0.32\linewidth}
          \centering
          \includegraphics[width=\textwidth]{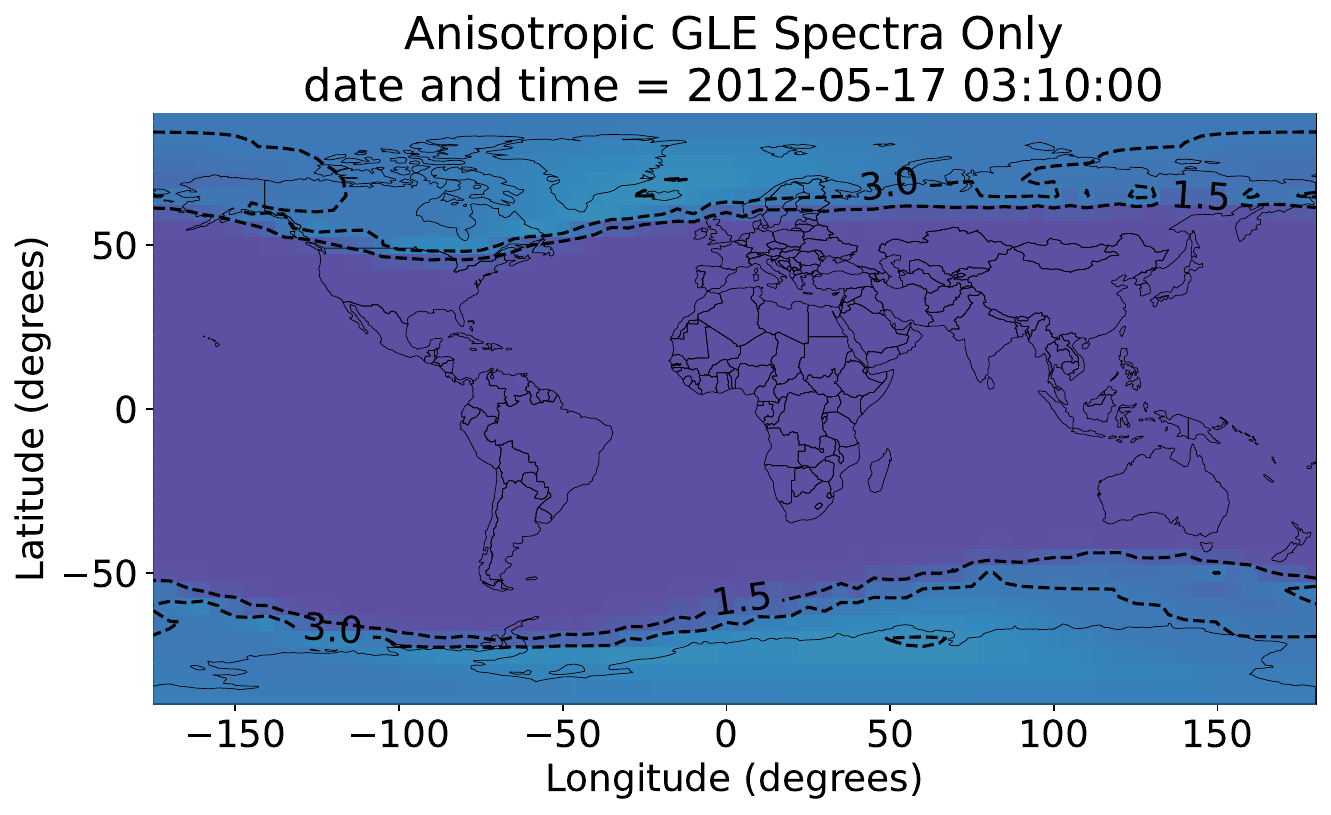}
          \caption{}
          \label{fig:sfig1}
    \end{subfigure}
    \begin{subfigure}[t]{0.32\linewidth}
          \centering
          \includegraphics[width=\textwidth]{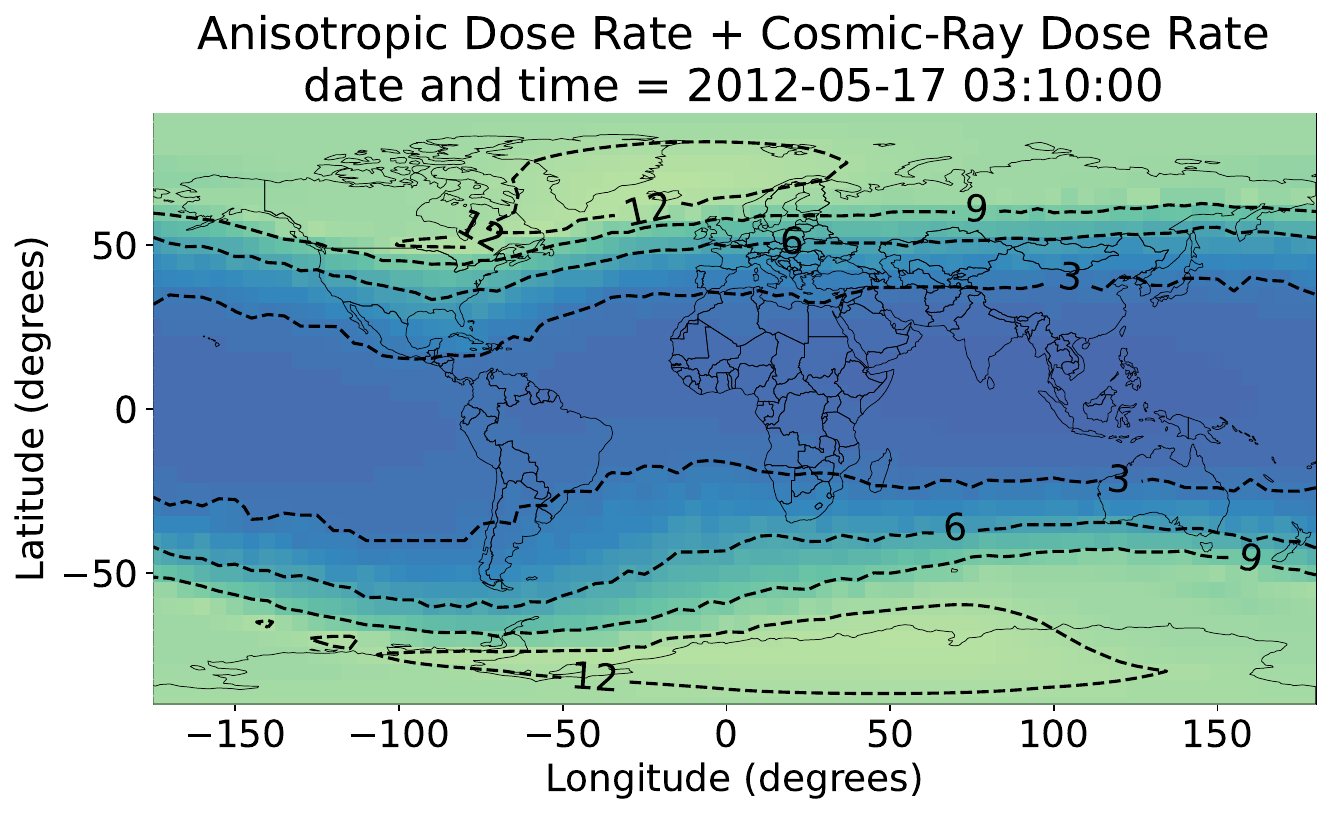}
          \caption{}
          \label{fig:sfig1}
    \end{subfigure}

    \begin{subfigure}[t]{\linewidth}
          \centering
          \includegraphics[width=\textwidth,trim={0 0 0 0.6\textheight}, clip]{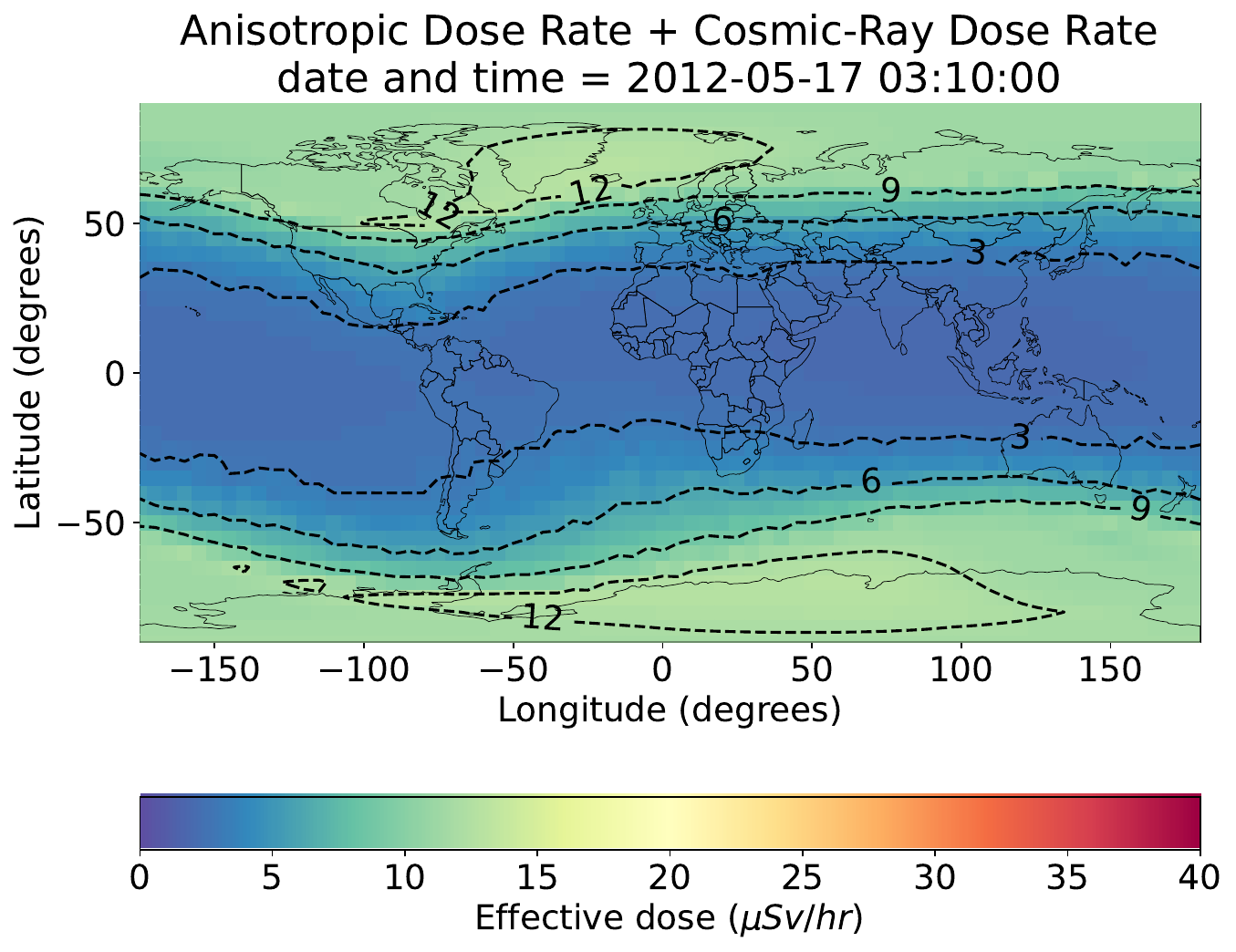}
    \end{subfigure}
    \caption{Effective dose maps for an altitude of 39~kft for an isotropic GLE71 spectrum, the anisotropic GLE71 spectrum and pitch angle distribution, and the total (anisotropic GLE + cosmic ray) dose for the anisotropic GLE71. As the isotropic spectrum is a worst-case assumption, global dose rates are overall significantly larger across Earth in the isotropic case than in the anisotropic case but are missing any of the complex anisotropic structures that are present in the anisotropic case.}
    \label{fig:GLEdoseMaps}
\end{figure}

\Cref{fig:GLEdoseMaps} displays three dose rate maps calculated at three selected time periods during the event. The three maps calculated for each time period were: an isotropic map, an anisotropic dose rate map, and a total dose rate map, which combines the anisotropic dose rate map with a background, cosmic ray dose rate map. An animated map showing dose rates throughout the whole event has been attached in the supplemental material for this paper, as well as outputted dose rate data files.

The anisotropic dose rate maps were calculated as described above. For the isotropic dose rate maps, $G(\alpha(P))$ was set to 1 for all pitch angles, $\alpha$. This demonstrates the type of result which can be achieved with isotropic-only models (which is a worst-case scenario), using the method of Mishev et al. for isotropic dose calculation. This method is also similar to that of \citeA{copeland2008cosmic}.  For total dose rate maps, the anisotropic dose rate map was added to the cosmic ray dose map, which in this case was calculated using the previously mentioned DLR-modified ISO model  \cite{matthia2013ready} as implemented in the CosRayModifiedISO Python module  \cite{cosraymodifiediso}, supplied with a pre-event OULU neutron monitor count rate of 106.54 cts/s. 

\Cref{fig:GLEdoseMaps} shows large differences between the isotropic and anisotropic situations at the beginning of the event. The isotropic situation shows broadly the same dose rates across the polar regions beyond a transition region. In contrast, for the anisotropic event, there are two "peaks" in the polar regions, corresponding to the dual peaks of the double Gaussian pitch angle distribution, but with a complex geographic curved shape formed due to the influence of Earth's magnetosphere and asymptotic directions. As the event proceeds over time, the anisotropic maps become steadily more similar to the isotropic maps due to the spreading of the Gaussian widths for the pitch angle distribution. This convergence over time towards a more isotropic distribution is common in GLEs \cite{shea2012space}. However, noticeable differences in dose rate maps can be observed for this event even 65 minutes after it began.

\begin{figure}
    \centering
    \includegraphics[width=\textwidth]{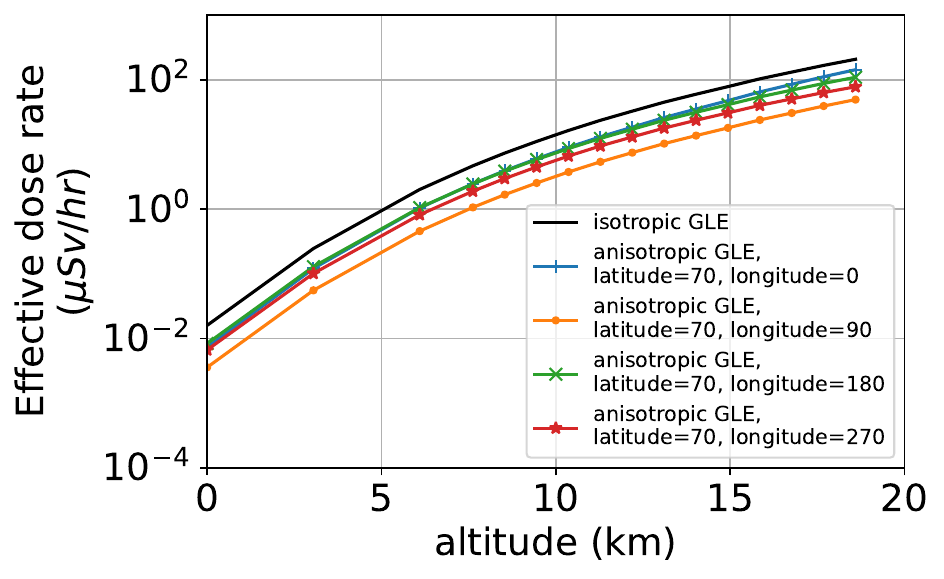}
    \caption{Effective dose rates as a function of altitude at different polar latitude-longitude locations. The isotropic distributions are larger in all cases, which is expected, and the curves are similarly shaped but do vary in curvature, particularly towards higher altitudes.}
    \label{fig:altitudeDependence}
\end{figure}

In addition to investigating dose rates as a function of location on Earth, the dependence of dose rates on altitude in the isotropic case vs the anisotropic case has been examined. The altitude dependence of dose rates is shown in \cref{fig:altitudeDependence} for the isotropic case in the polar region and at four polar coordinates surrounding Earth in the anisotropic case. 

While it is theoretically possible that the anisotropy could cause the effective dose-altitude curve to vary in shape, the curves in \cref{fig:altitudeDependence} all have similar shapes. There is nevertheless some divergence in shape, particularly at higher altitudes. For instance, the longitude=0.0~\textdegree~  and longitude=180.0~\textdegree~  cases begin to diverge at about 13~km in altitude.

\begin{figure}
    \centering
    \begin{subfigure}[b]{0.75\linewidth}
          \centering
          \includegraphics[width=\textwidth,trim={0 3cm 0 4cm},clip]{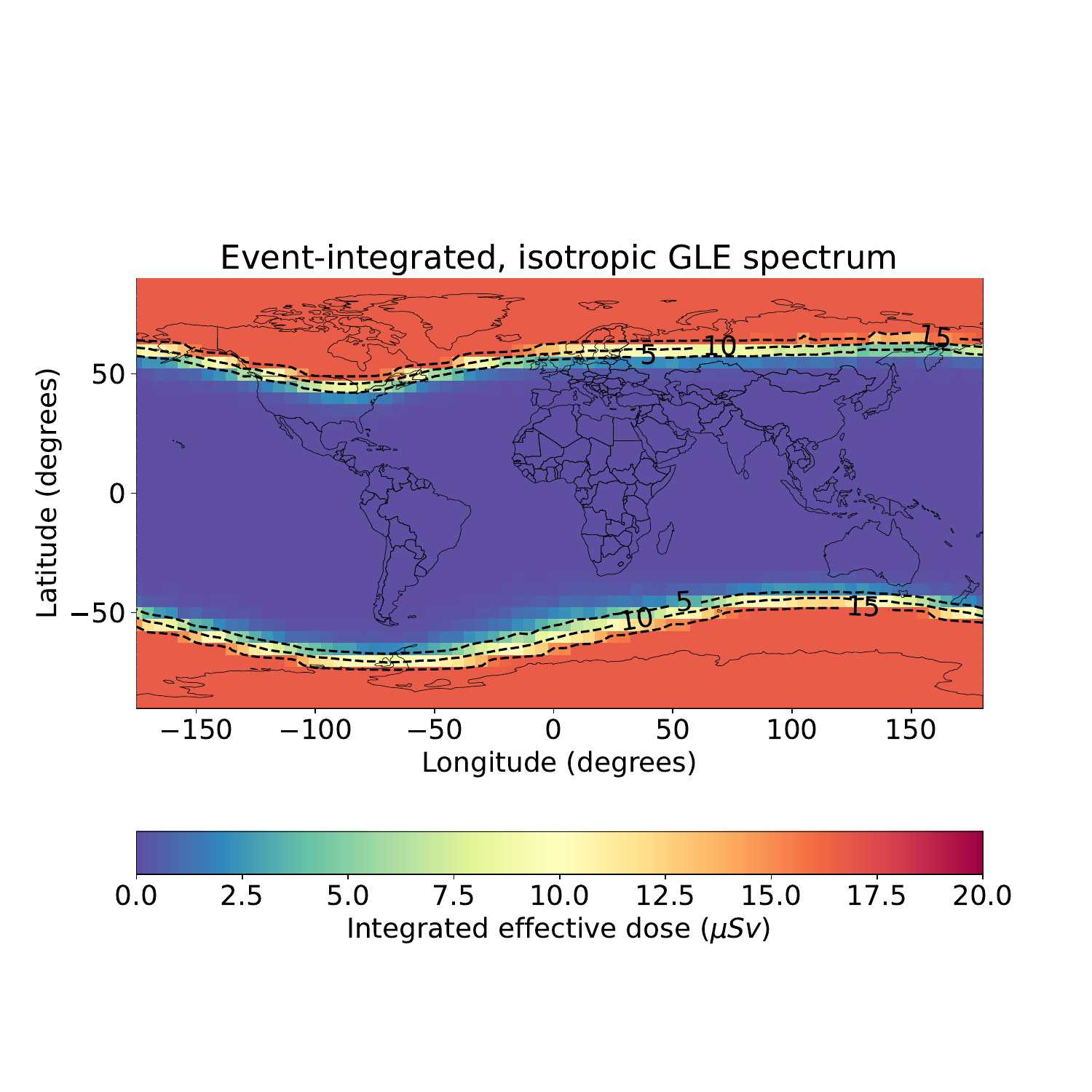}
          \caption{}
          \label{fig:isotropicIntegrated}
    \end{subfigure}
    \begin{subfigure}[b]{0.75\linewidth}
          \centering
          \includegraphics[width=\textwidth,trim={0 3cm 0 4cm},clip]{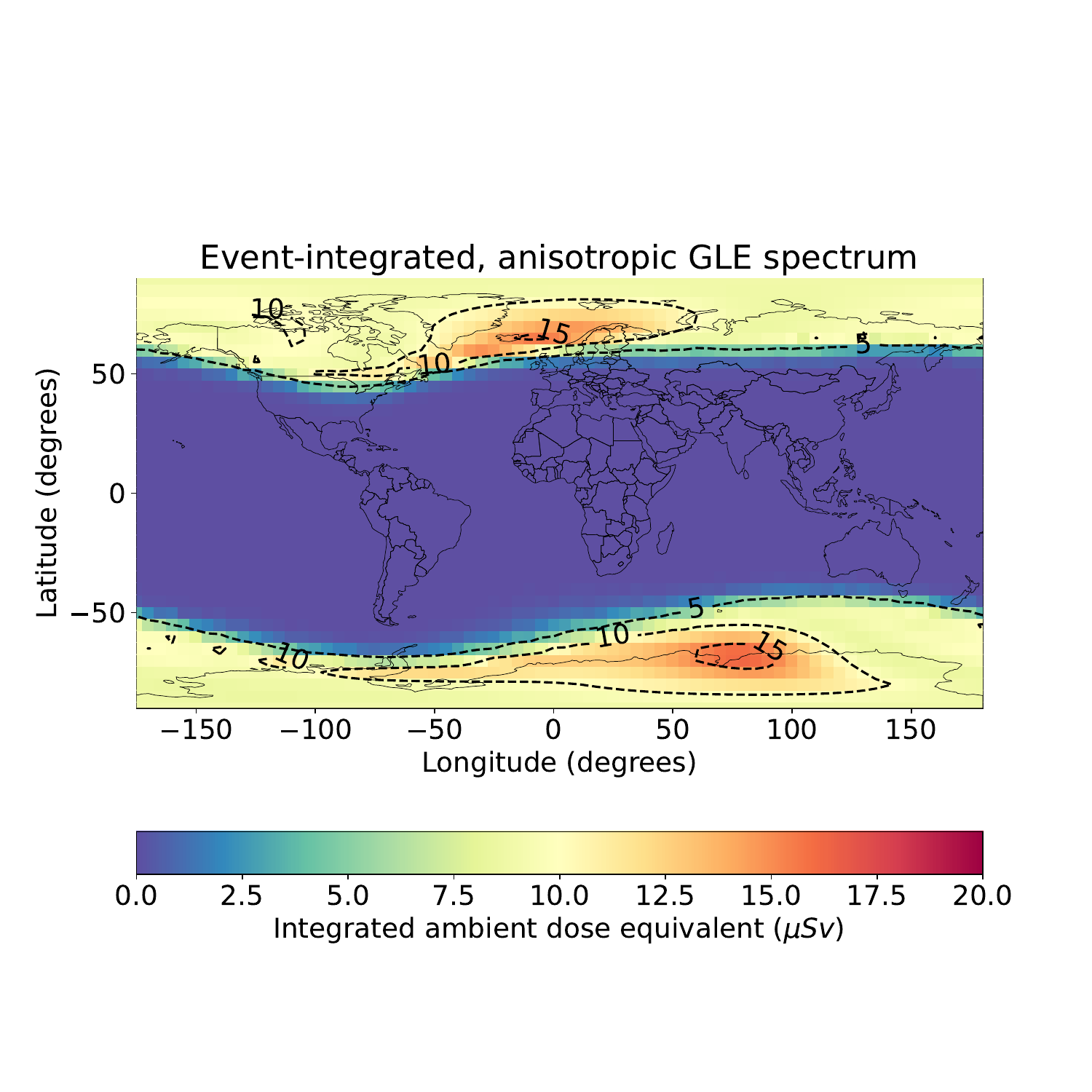}
          \caption{}
          \label{fig:anisotropicIntegrated}
    \end{subfigure}
    \caption{Integrated dose maps across the full event as defined by the parameters given by Mishev et al., which were fed into AniMAIRE for both the isotropic case and anisotropic case at 39~kft (excluding the final 5 minute timestamp due to a minor numerical issue with applying parameters to the supplied spectrum equation). Although the event is mostly anisotropic during only the first half an hour, the anisotropy can clearly be observed in the anisotropic map. This illustrates how the anisotropy must be considered to accurately assess the total dose a flight will receive, even if the anisotropy is primarily dominant specifically at the beginning of an event.}
    \label{fig:integrated_dose_rate_maps}
\end{figure}

One question of interest that arises is whether or not anisotropy in the event is important for event-integrated doses, as anisotropy is typically only significant in the beginning stages of an event. We find that for this event, the anisotropy is indeed important, and this can be seen in the event-integrated dose maps shown in \cref{fig:integrated_dose_rate_maps}.

Wide hotspots of high dose rates in the polar regions can be observed in \cref{fig:anisotropicIntegrated}, in contrast to the relatively flat polar dose rates in the isotropic case. This indicates that the total dose experienced by an aircraft cannot be calculated accurately using only isotropic assumptions - any aircraft positioned in one of the polar anisotropic hot regions at the beginning of the event will accumulate a much larger dose than an aircraft positioned at a different polar region.

\section{Comparisons to other models}

\begin{figure}
    \centering
    \begin{subfigure}[b]{0.45\linewidth}
          \centering
          \includegraphics[width=\textwidth]{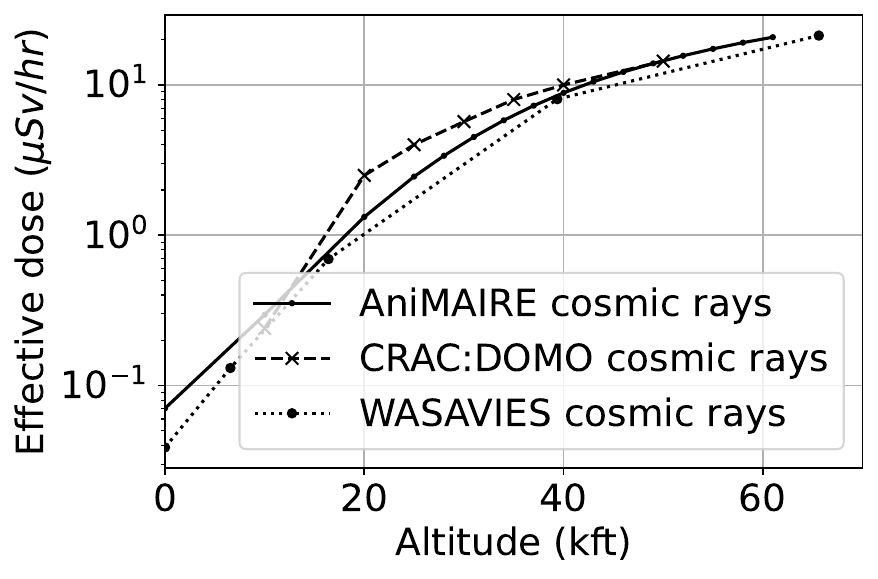}
          \caption{}
          \label{fig:cosRayComparison}
    \end{subfigure}
    \begin{subfigure}[b]{0.45\linewidth}
          \centering
          \includegraphics[width=\textwidth]{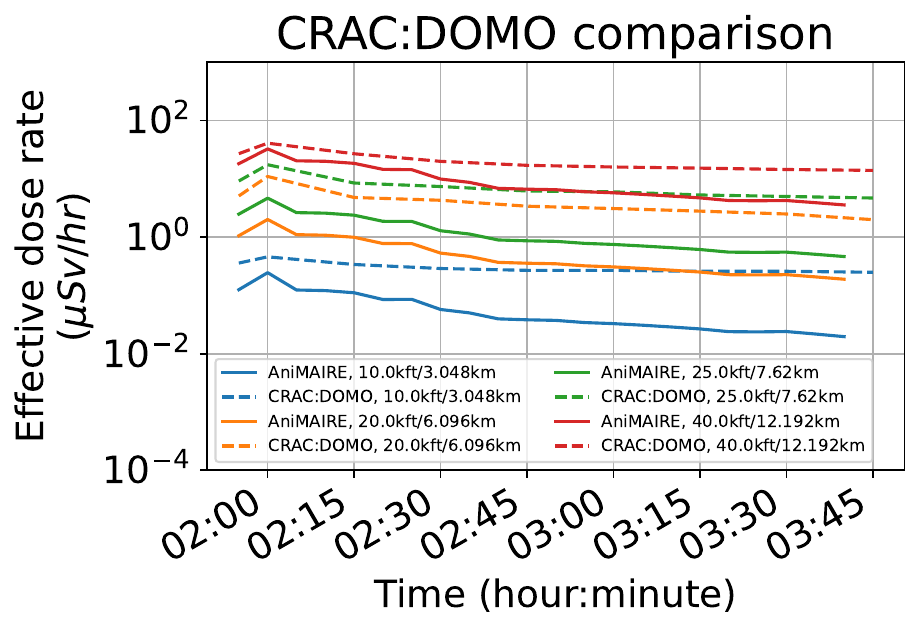}
          \caption{}
          \label{fig:MishevComparison}
    \end{subfigure}
    \begin{subfigure}[b]{0.45\linewidth}
          \centering
          \includegraphics[width=\textwidth]{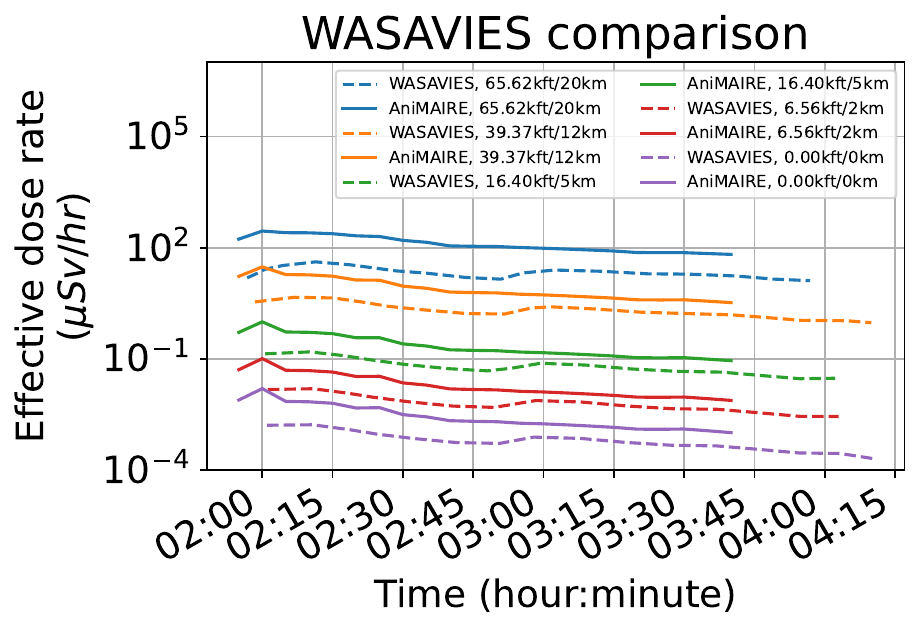}
          \caption{}
          \label{fig:WASAVIESComparison}
    \end{subfigure}
    \begin{subfigure}[b]{0.45\linewidth}
          \centering
          \includegraphics[width=\textwidth]{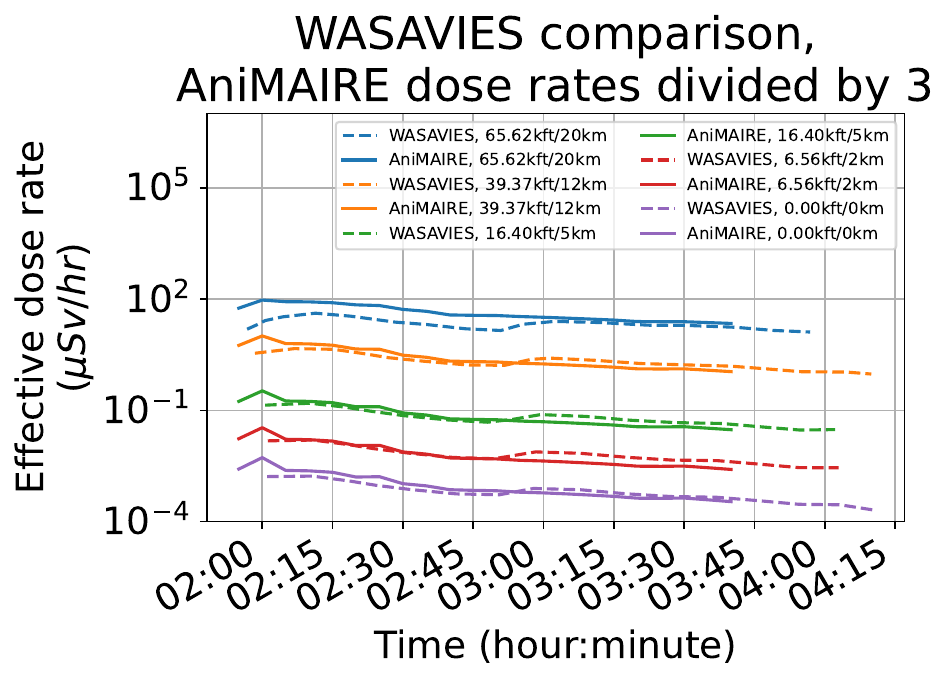}
          \caption{}
          \label{fig:WASAVIESComparisonDividedBy3}
    \end{subfigure}
    \caption{Comparisons with maximum dose results from CRAC:DOMO and WASAVIES. Comparisons between cosmic ray dose rates during GLE71 can be seen in \cref{fig:cosRayComparison}. Comparisons with GLE-only effective dose rates from CRAC:DOMO are shown in \cref{fig:MishevComparison}, and comparisons with WASAVIES data are shown in \cref{fig:WASAVIESComparison,fig:WASAVIESComparisonDividedBy3}, where AniMAIRE dose rates in \cref{fig:WASAVIESComparisonDividedBy3} have been arbitrarily divided by a factor of 3 to show how the relationship between effective dose rates calculated by AniMAIRE and WASAVIES match extremely well, except for differing by a flat factor, indicating that this difference likely arises from a difference in input spectra normalisation.}
    \label{fig:isotropicValidation}
\end{figure}

In addition to analysing the event, it was also possible to compare isotropic AniMAIRE GLE71 effective dose rates to dose rates produced by CRAC:DOMO and WASAVIES. \Cref{fig:cosRayComparison} shows isotropic GCR dose rates as a function of altitude compared with dose rates given by CRAC:DOMO \cite{mishev2021application} and WASAVIES \cite{sato2018real}. Comparisons with the GLE effective dose rates from CRAC:DOMO are shown in \cref{fig:MishevComparison}, and comparisons with WASAVIES data are shown in \cref{fig:WASAVIESComparison,fig:WASAVIESComparisonDividedBy3}.

The GCR dose rates given by CRAC:DOMO are consistently larger than those produced by AniMAIRE, particularly in the 20~kft to 40~kft range. The GLE effective dose rates of CRAC:DOMO over time also follow a different profile to those produced by AniMAIRE, as the dose rates of CRAC:DOMO decrease at a much slower rate over time. Interestingly, the opposite conclusions are true for the comparisons with WASAVIES dose rates. WASAVIES dose rates are significantly lower than those produced by AniMAIRE; however, the variation in dose rates over time matches extremely well, as does the ratio between dose rates at different altitudes. This is illustrated in \cref{fig:WASAVIESComparison,fig:WASAVIESComparisonDividedBy3} where the AniMAIRE dose vs time curves match the WASAVIES curves remarkably well when divided flatly by an arbitrarily chosen value of three.

A plausible explanation for the differences in shape and structure between AniMAIRE and CRAC:DOMO is provided by the plot of cosmic ray comparisons shown in \cref{fig:cosRayComparison}. AniMAIRE's cosmic ray-induced dose rates agree well with the WASAVIES cosmic ray-induced dose rates above 15~kft. In contrast, the CRAC:DOMO cosmic ray-induced dose rates are much larger than the AniMAIRE dose rates between about 20~kft and 35~kft. This implies that AniMAIRE and WASAVIES use similar yield functions to calculate dose rates from the incoming spectra but that the yield functions used by CRAC:DOMO are much larger in the 20~kft to 35~kft region. This explains why the differences between CRAC:DOMO and AniMAIRE's dose rates in \cref{fig:MishevComparison} are largest in the 20~kft to 35~kft region and it also explains why the AniMAIRE dose rate curves match the shape and structure of WASAVIES but not CRAC:DOMO's.

The fact that dose rates produced by AniMAIRE in this instance are significantly larger than those predicted by WASAVIES is unsurprising as a previous study by \citeA{sato2018real} using WASAVIES on GLE69 found that the effective dose rates produced were a factor of between 2 to 3 lower than prior published dose rates, including a factor of 2 lower than Mishev et al.'s model (now named CRAC:DOMO) for that GLE. Sato et al. explain this as WASAVIES predicting lower spectral indices for some GLEs than many other models, which also causes other parameters in WASAVIES to vary to compensate, leading to lower predicted dose rates. Note that WASAVIES here calculated spectral indices to a precision of 0.5, and that a spectral index of exactly 4.5 was used for almost the entirety of the first hour of GLE71.


Overall, these comparisons show that the AniMAIRE results for GLE71 fall between those of CRAC:DOMO and WASAVIES, but agree better with WASAVIES in terms of altitude structure and time variations. In future work, it would be interesting for AniMAIRE to be run with other spectra and pitch angle distributions and compared to other anisotropic models to see if these differences remain, and it would also be interesting for comparisons to be performed using other spectrum calculation algorithms and magnetospheric models, which could be added to future versions of AniMAIRE.

\section{Conclusion\remove{,} and Future Possible Extensions}

AniMAIRE is a new open-source toolkit for calculating radiation dose rates due to GLEs and cosmic rays, and the openly available nature of AniMAIRE should provide the community with an easy way to investigate how dose rates on Earth vary during anisotropic conditions. 

The analysis of the GLE71 dose rates calculated by AniMAIRE described in this paper found several interesting results and showcases the capabilities of the model. Comparing the dose rates calculated with assumptions of isotropy and anisotropy highlights the variation in predicted radiation exposure between the two methods. While this difference is most notable at early times, comparison of event-integrated dose rate maps shows that the proper treatment of the anisotropy in SEP protons is key to accurately predicting dose rates during GLEs.

The dose rates predicted by AniMAIRE for GLE71 have also been compared to those predicted by WASAVIES and Mishev et al. (CRAC:DOMO). The time-profile of dose rates calculated with AniMAIRE agrees well with WASAVIES, although the dose rates predicted by AniMAIRE are 3-4 times higher than those predicted by WASAVIES. This discrepancy is consistent with previous comparisons which found WASAVIES predicted lower counting rates than other models.  In contrast, the dose rates predicted by CRAC:DOMO are higher than those predicted by AniMAIRE and decay much more slowly in time. This may be explained in part by CRAC:DOMO employing yield functions which result in a much higher dose at aviation altitudes.

The Pythonic design of AniMAIRE should mean it is relatively easy to plug into other projects, and other members of the space weather community could connect AniMAIRE with their own tools. Some ideas that could be implemented in the future include adding GLE spectral calculations to AniMAIRE, perhaps for instance using the MAIRE+ GLE spectrum algorithm  \cite{hands2022new}. Another idea might be to add options for using magnetospheric models other than Tsyganenko 1989 or even alternatives to MAGNETOCOSMICS such as OTSO \cite{larsen2023new}. Finally, work is ongoing in collaboration with the Community Coordinated Modeling Center (CCMC) \cite{CCMC} at NASA and with ESA through their Network of Models (NoM) \cite{NoM} to create web interfaces for AniMAIRE so non-Python users can perform runs without prior programming experience.

\section{Software and Data Availability}
\label{sec:DataAvailability}

AniMAIRE, and all of the software that have been developed as part of this project have been released openly online. The AniMAIRE code itself is available at its homepage on Github at \url{https://github.com/ssc-maire/AniMAIRE-public} . The README file in the Github repository contains information on both how to install AniMAIRE, and how to use it, including example commands and outputs. The Github repository also contains two Jupyter Notebooks called \url{AniMAIRE_examples.ipynb}, which can be run by users to produce several example dose rate maps, and \url{notebooks_and_data_and_figures_for_paper/GLE71_plots_for_paper.ipynb}, which was the Jupyter Notebook used to generate all of the plots and data used for this paper. All of the data and figures used for this paper can also be found in the \url{notebooks_and_data_and_figures_for_paper/} directory and subdirectories in the Github repository.

While AniMAIRE can be installed from Github directly, we would recommend users actually install it using the command:
\begin{verbatim}
      pip install AniMAIRE
\end{verbatim}
in a Linux terminal. This command installs AniMAIRE from its PyPi page at \url{https://pypi.org/project/AniMAIRE/} and almost every package it depends on automatically, without the need for any further commands. Note that if you use Windows rather than Linux, you can easily install Windows Subsystem for Linux (WSL) from \url{https://apps.microsoft.com/detail/9mttcl66cpxj}, which is a Linux terminal that runs on Windows systems (AniMAIRE was actually developed using WSL and has been tested on the Ubuntu 20.04.6 LTS distribution of WSL).

The only additional package you need to additionally manually install at present to run AniMAIRE is the Geant4-based MAGNETOCOSMICS package \cite{MAGNETOCOSMICS,desorgher2004magnetocosmics} (the AsympDirsCalculator package, which AniMAIRE currently uses, requires MAGNETOCOSMICS to calculate asymptotic directions and cut-off rigidities). MAGNETOCOSMICS can be found and installed from \url{https://cosray.unibe.ch/~laurent/magnetocosmics/}, and also from several other locations around the internet. If you are struggling to install MAGNETOCOSMICS or have any difficulties or questions in general regarding AniMAIRE installation and usage, please feel free to contact us through posting on the discussion section at the AniMAIRE Github page, direct Github messaging, or by emailing us. 

Some users may also be interested in separately using some of the Python packages AniMAIRE depends on (and which get automatically installed with AniMAIRE) that we developed as part of this project, some of which we've found very useful on their own for our research. These are:

\begin{enumerate}
      \item \textit{ParticleRigidityCalculationTools} \cite{ParticleRigidityCalculationTools}: This is a Python package for performing simple conversions between kinetic energy and rigidity, and kinetic energy spectra and rigidity spectra. Its Github repository is at \url{https://github.com/ChrisSWDavis/ParticleRigidityCalculationTools} and its installation page is at \url{https://pypi.org/project/ParticleRigidityCalculationTools/} . Its Zenodo DOI is \url{https://zenodo.org/doi/10.5281/zenodo.10992232} .
      \item \textit{DoseAndFluxCalculator} \cite{atmosphericRadiationDoseAndFlux}: This is a Python package for calculating dose rates and single event effects due to a particular given particle spectrum hitting Earth's atmosphere, by applying the yield functions from previously developed MAIRE models \cite{MAIREUK,hands2022new} to the inputted spectra. Its Github repository is at \url{https://github.com/ssc-maire/DoseAndFluxCalculator} and its installation page is at \url{https://pypi.org/project/atmosphericRadiationDoseAndFlux/} . Its Zenodo DOI is \url{https://zenodo.org/doi/10.5281/zenodo.10992476} .
      \item \textit{CosRayModifiedISO} \cite{cosraymodifiediso}: This is a Python package for calculating spectra due to cosmic rays given a particular date and time, solar modulation or OULU neutron monitor count rate. It is based on a model developed by \citeA{matthia2013ready}. Its Github repository is at \url{https://github.com/ssc-maire/CosRayModifiedISO} and its installation page is at \url{https://pypi.org/project/CosRayModifiedISO/} . Its Zenodo DOI is \url{https://zenodo.org/doi/10.5281/zenodo.10992395} .
      \item \textit{AsymptoticDirectionsCalculator} \cite{AsympDirsCalculator}: This is a Python package for calculating asymptotic directions and vertical cut-off rigidities. Currently it effectively acts as a Python wrapper for running MAGNETOCOSMICS, although there have been discussions to eventually change this to an optional dependency, and perhaps add alternative models such as OTSO\cite{larsen2023new}. Its Github repository is at \url{https://github.com/ssc-maire/AsymptoticDirectionsCalculator-public} and its installation page is at \url{https://pypi.org/project/AsympDirsCalculator/} . Its Zenodo DOI is \url{https://zenodo.org/doi/10.5281/zenodo.10992606} .
\end{enumerate}

Several animations can also be found in the supplemental materials of this paper and in the AniMAIRE Github repository, which show the full progression of GLE71 maps throughout the whole event. All of AniMAIRE's releases, its data, and plots have been archived on Zenodo at the DOI \url{https://zenodo.org/doi/10.5281/zenodo.10992643}, and the licences for all bespoke software described in this paper can be found in the software repositories.







\acknowledgments

We would like to thank Alexander Mishev, Nicholas Larsen and Sergey Koldobskiy for their help and advice while performing this research, and aiding us with performing comparisons with their work. 

We also acknowledge the NMDB database (www.nmdb.eu), founded under the European Union's FP7 programme (contract no. 213007) for providing data.

The SWARM project, that this software's development was part of, was funded by NERC (grant no. NE/V002899/1) under the SWIMMR programme, under the Strategic Priorities Fund (SPF) delivered by UKRI.


\bibliography{agusample}

\begin{thebibliography}{}

\bibitem [\protect \citeauthoryear {%
Bender%
}{%
Bender%
}{%
{\protect \APACyear {{\protect \bibnodate {}}}}%
}]{%
PySpaceWeather}
\APACinsertmetastar {%
PySpaceWeather}%
\begin{APACrefauthors}%
Bender, S.%
\end{APACrefauthors}%
\unskip\
\newblock
\APACrefYearMonthDay{{\protect \bibnodate {}}}{}{}.
\newblock
{\BBOQ}\APACrefatitle {{PySpaceWeather}} {{PySpaceWeather}}{\BBCQ}\
  [\bibcomputersoftware].
\newblock
\begin{APACrefURL} \url{https://github.com/st-bender/pyspaceweather}
  \end{APACrefURL}
\newblock
\APACrefnote{version 0.2.4}
\PrintBackRefs{\CurrentBib}

\bibitem [\protect \citeauthoryear {%
Bruno%
\ \protect \BOthers {.}}{%
Bruno%
\ \protect \BOthers {.}}{%
{\protect \APACyear {2018}}%
}]{%
bruno2018solar}
\APACinsertmetastar {%
bruno2018solar}%
\begin{APACrefauthors}%
Bruno, A.%
, Bazilevskaya, G.%
, Boezio, M.%
, Christian, E\BPBI R.%
, De~Nolfo, G.%
, Martucci, M.%
\BDBL {}others%
\end{APACrefauthors}%
\unskip\
\newblock
\APACrefYearMonthDay{2018}{}{}.
\newblock
{\BBOQ}\APACrefatitle {Solar energetic particle events observed by the PAMELA
  mission} {Solar energetic particle events observed by the pamela
  mission}.{\BBCQ}
\newblock
\APACjournalVolNumPages{The Astrophysical Journal}{862}{2}{97}.
\PrintBackRefs{\CurrentBib}

\bibitem [\protect \citeauthoryear {%
Cannon%
\ \protect \BOthers {.}}{%
Cannon%
\ \protect \BOthers {.}}{%
{\protect \APACyear {2013}}%
}]{%
cannon2013extreme}
\APACinsertmetastar {%
cannon2013extreme}%
\begin{APACrefauthors}%
Cannon, P.%
, Angling, M.%
, Barclay, L.%
, Curry, C.%
, Dyer, C.%
, Edwards, R.%
\BDBL {}others%
\end{APACrefauthors}%
\unskip\
\newblock
\APACrefYear{2013}.
\newblock
\APACrefbtitle {Extreme space weather: impacts on engineered systems and
  infrastructure} {Extreme space weather: impacts on engineered systems and
  infrastructure}.
\newblock
\APACaddressPublisher{}{Royal Academy of Engineering}.
\PrintBackRefs{\CurrentBib}

\bibitem [\protect \citeauthoryear {%
{Community Coordinated Modelling Center}%
}{%
{Community Coordinated Modelling Center}%
}{%
{\protect \APACyear {{\protect \bibnodate {}}}}%
}]{%
CCMC}
\APACinsertmetastar {%
CCMC}%
\begin{APACrefauthors}%
{Community Coordinated Modelling Center}.%
\end{APACrefauthors}%
\unskip\
\newblock
\APACrefYearMonthDay{{\protect \bibnodate {}}}{}{}.
\newblock
\APACrefbtitle {{CCMC}.} {{CCMC}.}
\newblock
\APACrefnote{\url{https://ccmc.gsfc.nasa.gov}}
\PrintBackRefs{\CurrentBib}

\bibitem [\protect \citeauthoryear {%
Cooke%
\ \protect \BOthers {.}}{%
Cooke%
\ \protect \BOthers {.}}{%
{\protect \APACyear {1991}}%
}]{%
Cooke1991}
\APACinsertmetastar {%
Cooke1991}%
\begin{APACrefauthors}%
Cooke, D\BPBI J.%
, Humble, J\BPBI E.%
, Shea, M\BPBI A.%
, Smart, D\BPBI F.%
, Lund, N.%
, Rasmussen, I\BPBI L.%
\BDBL {}Petrou, N.%
\end{APACrefauthors}%
\unskip\
\newblock
\APACrefYearMonthDay{1991}{5}{}.
\newblock
{\BBOQ}\APACrefatitle {On cosmic-ray cut-off terminology} {On cosmic-ray
  cut-off terminology}.{\BBCQ}
\newblock
\APACjournalVolNumPages{Il Nuovo Cimento C}{14}{}{213-234}.
\newblock
\begin{APACrefURL} \url{https://link.springer.com/article/10.1007/BF02509357}
  \end{APACrefURL}
\newblock
\begin{APACrefDOI} \doi{10.1007/BF02509357/METRICS} \end{APACrefDOI}
\PrintBackRefs{\CurrentBib}

\bibitem [\protect \citeauthoryear {%
Copeland%
}{%
Copeland%
}{%
{\protect \APACyear {2017}}%
}]{%
copeland2017cari}
\APACinsertmetastar {%
copeland2017cari}%
\begin{APACrefauthors}%
Copeland, K.%
\end{APACrefauthors}%
\unskip\
\newblock
\APACrefYearMonthDay{2017}{}{}.
\newblock
{\BBOQ}\APACrefatitle {CARI-7A: Development and validation} {Cari-7a:
  Development and validation}.{\BBCQ}
\newblock
\APACjournalVolNumPages{Radiation Protection Dosimetry}{175}{4}{419--431}.
\PrintBackRefs{\CurrentBib}

\bibitem [\protect \citeauthoryear {%
Copeland%
, Sauer%
, Duke%
\BCBL {}\ \BBA {} Friedberg%
}{%
Copeland%
\ \protect \BOthers {.}}{%
{\protect \APACyear {2008}}%
}]{%
copeland2008cosmic}
\APACinsertmetastar {%
copeland2008cosmic}%
\begin{APACrefauthors}%
Copeland, K.%
, Sauer, H\BPBI H.%
, Duke, F\BPBI E.%
\BCBL {}\ \BBA {} Friedberg, W.%
\end{APACrefauthors}%
\unskip\
\newblock
\APACrefYearMonthDay{2008}{}{}.
\newblock
{\BBOQ}\APACrefatitle {Cosmic radiation exposure of aircraft occupants on
  simulated high-latitude flights during solar proton events from 1 January
  1986 through 1 January 2008} {Cosmic radiation exposure of aircraft occupants
  on simulated high-latitude flights during solar proton events from 1 january
  1986 through 1 january 2008}.{\BBCQ}
\newblock
\APACjournalVolNumPages{Advances in Space Research}{42}{6}{1008--1029}.
\PrintBackRefs{\CurrentBib}

\bibitem [\protect \citeauthoryear {%
Cramp%
\ \protect \BOthers {.}}{%
Cramp%
\ \protect \BOthers {.}}{%
{\protect \APACyear {1997}}%
}]{%
cramp1997october}
\APACinsertmetastar {%
cramp1997october}%
\begin{APACrefauthors}%
Cramp, J.%
, Duldig, M.%
, Fl{\"u}ckiger, E.%
, Humble, J.%
, Shea, M.%
\BCBL {}\ \BBA {} Smart, D.%
\end{APACrefauthors}%
\unskip\
\newblock
\APACrefYearMonthDay{1997}{}{}.
\newblock
{\BBOQ}\APACrefatitle {The October 22, 1989, solar cosmic ray enhancement: An
  analysis of the anisotropy and spectral characteristics} {The october 22,
  1989, solar cosmic ray enhancement: An analysis of the anisotropy and
  spectral characteristics}.{\BBCQ}
\newblock
\APACjournalVolNumPages{Journal of Geophysical Research: Space
  Physics}{102}{A11}{24237--24248}.
\PrintBackRefs{\CurrentBib}

\bibitem [\protect \citeauthoryear {%
Davis%
}{%
Davis%
}{%
{\protect \APACyear {2024}}%
{\protect \APACexlab {{\protect \BCnt {1}}}}}]{%
AsympDirsCalculator}
\APACinsertmetastar {%
AsympDirsCalculator}%
\begin{APACrefauthors}%
Davis, C\BPBI S\BPBI W.%
\end{APACrefauthors}%
\unskip\
\newblock
\APACrefYearMonthDay{2024{\protect \BCnt {1}}}{{\APACmonth{04}}}{}.
\newblock
{\BBOQ}\APACrefatitle {{AsymptoticDirectionsCalculator-public}}
  {{AsymptoticDirectionsCalculator-public}}{\BBCQ}\ [\bibcomputersoftware].
\newblock
\begin{APACrefURL}
  \url{https://github.com/ssc-maire/AsymptoticDirectionsCalculator-public}
  \end{APACrefURL}
\newblock
\APACrefnote{version 1.0.9}
\newblock
\begin{APACrefDOI} \doi{10.5281/zenodo.10992606} \end{APACrefDOI}
\PrintBackRefs{\CurrentBib}

\bibitem [\protect \citeauthoryear {%
Davis%
}{%
Davis%
}{%
{\protect \APACyear {2024}}%
{\protect \APACexlab {{\protect \BCnt {2}}}}}]{%
cosraymodifiediso}
\APACinsertmetastar {%
cosraymodifiediso}%
\begin{APACrefauthors}%
Davis, C\BPBI S\BPBI W.%
\end{APACrefauthors}%
\unskip\
\newblock
\APACrefYearMonthDay{2024{\protect \BCnt {2}}}{{\APACmonth{04}}}{}.
\newblock
{\BBOQ}\APACrefatitle {{CosRayModifiedISO}} {{CosRayModifiedISO}}{\BBCQ}\
  [\bibcomputersoftware].
\newblock
\begin{APACrefURL} \url{https://github.com/ssc-maire/CosRayModifiedISO}
  \end{APACrefURL}
\newblock
\APACrefnote{version 1.2.4}
\newblock
\begin{APACrefDOI} \doi{10.5281/zenodo.10992395} \end{APACrefDOI}
\PrintBackRefs{\CurrentBib}

\bibitem [\protect \citeauthoryear {%
Davis%
}{%
Davis%
}{%
{\protect \APACyear {2024}}%
{\protect \APACexlab {{\protect \BCnt {3}}}}}]{%
ParticleRigidityCalculationTools}
\APACinsertmetastar {%
ParticleRigidityCalculationTools}%
\begin{APACrefauthors}%
Davis, C\BPBI S\BPBI W.%
\end{APACrefauthors}%
\unskip\
\newblock
\APACrefYearMonthDay{2024{\protect \BCnt {3}}}{{\APACmonth{04}}}{}.
\newblock
{\BBOQ}\APACrefatitle {{ParticleRigidityCalculationTools}}
  {{ParticleRigidityCalculationTools}}{\BBCQ}\ [\bibcomputersoftware].
\newblock
\begin{APACrefURL}
  \url{https://github.com/ChrisSWDavis/ParticleRigidityCalculationTools}
  \end{APACrefURL}
\newblock
\APACrefnote{version 1.5.7}
\newblock
\begin{APACrefDOI} \doi{10.5281/zenodo.10992232} \end{APACrefDOI}
\PrintBackRefs{\CurrentBib}

\bibitem [\protect \citeauthoryear {%
Davis%
\ \BBA {} Lei%
}{%
Davis%
\ \BBA {} Lei%
}{%
{\protect \APACyear {2024}}%
}]{%
atmosphericRadiationDoseAndFlux}
\APACinsertmetastar {%
atmosphericRadiationDoseAndFlux}%
\begin{APACrefauthors}%
Davis, C\BPBI S\BPBI W.%
\BCBT {}\ \BBA {} Lei, F.%
\end{APACrefauthors}%
\unskip\
\newblock
\APACrefYearMonthDay{2024}{{\APACmonth{04}}}{}.
\newblock
{\BBOQ}\APACrefatitle {{DoseAndFluxCalculator}}
  {{DoseAndFluxCalculator}}{\BBCQ}\ [\bibcomputersoftware].
\newblock
\begin{APACrefURL} \url{https://github.com/ssc-maire/DoseAndFluxCalculator}
  \end{APACrefURL}
\newblock
\APACrefnote{version 1.0.4}
\newblock
\begin{APACrefDOI} \doi{10.5281/zenodo.10992476} \end{APACrefDOI}
\PrintBackRefs{\CurrentBib}

\bibitem [\protect \citeauthoryear {%
Davis%
, Lei%
, Baird%
, Ryden%
\BCBL {}\ \BBA {} Dyer%
}{%
Davis%
\ \protect \BOthers {.}}{%
{\protect \APACyear {2024}}%
}]{%
AniMAIRE}
\APACinsertmetastar {%
AniMAIRE}%
\begin{APACrefauthors}%
Davis, C\BPBI S\BPBI W.%
, Lei, F.%
, Baird, F.%
, Ryden, K.%
\BCBL {}\ \BBA {} Dyer, C.%
\end{APACrefauthors}%
\unskip\
\newblock
\APACrefYearMonthDay{2024}{{\APACmonth{04}}}{}.
\newblock
{\BBOQ}\APACrefatitle {{AniMAIRE}} {{AniMAIRE}}{\BBCQ}\ [\bibcomputersoftware].
\newblock
\begin{APACrefURL} \url{https://github.com/ssc-maire/AniMAIRE-public}
  \end{APACrefURL}
\newblock
\APACrefnote{version 1.1.5}
\newblock
\begin{APACrefDOI} \doi{10.5281/zenodo.10992643} \end{APACrefDOI}
\PrintBackRefs{\CurrentBib}

\bibitem [\protect \citeauthoryear {%
Desorgher%
}{%
Desorgher%
}{%
{\protect \APACyear {{\protect \bibnodate {}}}}%
}]{%
PLANETOCOSMICS}
\APACinsertmetastar {%
PLANETOCOSMICS}%
\begin{APACrefauthors}%
Desorgher, L.%
\end{APACrefauthors}%
\unskip\
\newblock
\APACrefYearMonthDay{{\protect \bibnodate {}}}{}{}.
\newblock
\APACrefbtitle {PLANETOCOSMICS.} {Planetocosmics.}
\newblock
\begin{APACrefURL} \url{http://cosray.unibe.ch/∼laurent/planetocosmics/}
  \end{APACrefURL}
\PrintBackRefs{\CurrentBib}

\bibitem [\protect \citeauthoryear {%
Desorgher%
}{%
Desorgher%
}{%
{\protect \APACyear {2004}}%
}]{%
desorgher2004magnetocosmics}
\APACinsertmetastar {%
desorgher2004magnetocosmics}%
\begin{APACrefauthors}%
Desorgher, L.%
\end{APACrefauthors}%
\unskip\
\newblock
\APACrefYearMonthDay{2004}{}{}.
\newblock
{\BBOQ}\APACrefatitle {MAGNETOCOSMICS Software User Manual} {Magnetocosmics
  software user manual}{\BBCQ}\ [\bibcomputersoftwaremanual].
\newblock
\APACrefnote{Physikalisches Institut University of Bern, http://reat.space.
  qinetiq.com/septimess/magcos/magnetocosmics\_sum.pdf, Bern}
\PrintBackRefs{\CurrentBib}

\bibitem [\protect \citeauthoryear {%
Desorgher%
}{%
Desorgher%
}{%
{\protect \APACyear {2006}}%
}]{%
MAGNETOCOSMICS}
\APACinsertmetastar {%
MAGNETOCOSMICS}%
\begin{APACrefauthors}%
Desorgher, L.%
\end{APACrefauthors}%
\unskip\
\newblock
\APACrefYearMonthDay{2006}{{\APACmonth{09}}}{}.
\newblock
{\BBOQ}\APACrefatitle {{MAGNETOCOSMICS}} {{MAGNETOCOSMICS}}{\BBCQ}\
  [\bibcomputersoftware].
\newblock
\begin{APACrefURL} \url{http://cosray.unibe.ch/~laurent/magnetocosmics/}
  \end{APACrefURL}
\newblock
\APACrefnote{version 2.0}
\PrintBackRefs{\CurrentBib}

\bibitem [\protect \citeauthoryear {%
Desorgher%
, Fl{\"u}ckiger%
\BCBL {}\ \BBA {} Gurtner%
}{%
Desorgher%
\ \protect \BOthers {.}}{%
{\protect \APACyear {2006}}%
}]{%
desorgher2006planetocosmics}
\APACinsertmetastar {%
desorgher2006planetocosmics}%
\begin{APACrefauthors}%
Desorgher, L.%
, Fl{\"u}ckiger, E\BPBI O.%
\BCBL {}\ \BBA {} Gurtner, M.%
\end{APACrefauthors}%
\unskip\
\newblock
\APACrefYearMonthDay{2006}{}{}.
\newblock
{\BBOQ}\APACrefatitle {The planetocosmics geant4 application} {The
  planetocosmics geant4 application}.{\BBCQ}
\newblock
\BIn{} \APACrefbtitle {36th COSPAR Scientific Assembly} {36th cospar scientific
  assembly}\ (\BVOL~36, \BPG~2361).
\PrintBackRefs{\CurrentBib}

\bibitem [\protect \citeauthoryear {%
Dyer%
, Hands%
, Ryden%
\BCBL {}\ \BBA {} Lei%
}{%
Dyer%
\ \protect \BOthers {.}}{%
{\protect \APACyear {2017}}%
}]{%
dyer2017extreme}
\APACinsertmetastar {%
dyer2017extreme}%
\begin{APACrefauthors}%
Dyer, C.%
, Hands, A.%
, Ryden, K.%
\BCBL {}\ \BBA {} Lei, F.%
\end{APACrefauthors}%
\unskip\
\newblock
\APACrefYearMonthDay{2017}{}{}.
\newblock
{\BBOQ}\APACrefatitle {Extreme atmospheric radiation environments and single
  event effects} {Extreme atmospheric radiation environments and single event
  effects}.{\BBCQ}
\newblock
\APACjournalVolNumPages{IEEE Transactions on Nuclear Science}{65}{1}{432--438}.
\PrintBackRefs{\CurrentBib}

\bibitem [\protect \citeauthoryear {%
Dyer%
, Lei%
, Hands%
\BCBL {}\ \BBA {} Truscott%
}{%
Dyer%
\ \protect \BOthers {.}}{%
{\protect \APACyear {2007}}%
}]{%
dyer2007solar}
\APACinsertmetastar {%
dyer2007solar}%
\begin{APACrefauthors}%
Dyer, C.%
, Lei, F.%
, Hands, A.%
\BCBL {}\ \BBA {} Truscott, P.%
\end{APACrefauthors}%
\unskip\
\newblock
\APACrefYearMonthDay{2007}{}{}.
\newblock
{\BBOQ}\APACrefatitle {Solar particle events in the QinetiQ atmospheric
  radiation model} {Solar particle events in the qinetiq atmospheric radiation
  model}.{\BBCQ}
\newblock
\APACjournalVolNumPages{IEEE Transactions on Nuclear
  Science}{54}{4}{1071--1075}.
\PrintBackRefs{\CurrentBib}

\bibitem [\protect \citeauthoryear {%
Dyer%
, Sims%
, Farren%
\BCBL {}\ \BBA {} Stephen%
}{%
Dyer%
\ \protect \BOthers {.}}{%
{\protect \APACyear {1989}}%
}]{%
dyer1989measurements}
\APACinsertmetastar {%
dyer1989measurements}%
\begin{APACrefauthors}%
Dyer, C.%
, Sims, A.%
, Farren, J.%
\BCBL {}\ \BBA {} Stephen, J.%
\end{APACrefauthors}%
\unskip\
\newblock
\APACrefYearMonthDay{1989}{}{}.
\newblock
{\BBOQ}\APACrefatitle {Measurements of the SEU environment in the upper
  atmosphere} {Measurements of the seu environment in the upper
  atmosphere}.{\BBCQ}
\newblock
\APACjournalVolNumPages{IEEE Transactions on Nuclear
  Science}{36}{6}{2275--2280}.
\PrintBackRefs{\CurrentBib}

\bibitem [\protect \citeauthoryear {%
Dyer%
, Sims%
, Farren%
\BCBL {}\ \BBA {} Stephen%
}{%
Dyer%
\ \protect \BOthers {.}}{%
{\protect \APACyear {1990}}%
}]{%
dyer1990measurements}
\APACinsertmetastar {%
dyer1990measurements}%
\begin{APACrefauthors}%
Dyer, C.%
, Sims, A.%
, Farren, J.%
\BCBL {}\ \BBA {} Stephen, J.%
\end{APACrefauthors}%
\unskip\
\newblock
\APACrefYearMonthDay{1990}{}{}.
\newblock
{\BBOQ}\APACrefatitle {Measurements of solar flare enhancements to the single
  event upset environment in the upper atmosphere (avionics)} {Measurements of
  solar flare enhancements to the single event upset environment in the upper
  atmosphere (avionics)}.{\BBCQ}
\newblock
\APACjournalVolNumPages{IEEE Transactions on Nuclear
  Science}{37}{6}{1929--1937}.
\PrintBackRefs{\CurrentBib}

\bibitem [\protect \citeauthoryear {%
{European Space Agency}%
}{%
{European Space Agency}%
}{%
{\protect \APACyear {{\protect \bibnodate {}}}}%
}]{%
NoM}
\APACinsertmetastar {%
NoM}%
\begin{APACrefauthors}%
{European Space Agency}.%
\end{APACrefauthors}%
\unskip\
\newblock
\APACrefYearMonthDay{{\protect \bibnodate {}}}{}{}.
\newblock
\APACrefbtitle {{ESA Network of Models}.} {{ESA Network of Models}.}
\newblock
\APACrefnote{\url{https://nom.esa.int/}}
\PrintBackRefs{\CurrentBib}

\bibitem [\protect \citeauthoryear {%
Hands%
\ \protect \BOthers {.}}{%
Hands%
\ \protect \BOthers {.}}{%
{\protect \APACyear {2022}}%
}]{%
hands2022new}
\APACinsertmetastar {%
hands2022new}%
\begin{APACrefauthors}%
Hands, A.%
, Lei, F.%
, Davis, C.%
, Clewer, B.%
, Dyer, C.%
\BCBL {}\ \BBA {} Ryden, K.%
\end{APACrefauthors}%
\unskip\
\newblock
\APACrefYearMonthDay{2022}{}{}.
\newblock
{\BBOQ}\APACrefatitle {A New Model for Nowcasting the Aviation Radiation
  Environment With Comparisons to In Situ Measurements During GLEs} {A new
  model for nowcasting the aviation radiation environment with comparisons to
  in situ measurements during gles}.{\BBCQ}
\newblock
\APACjournalVolNumPages{Space Weather}{20}{8}{e2022SW003155}.
\PrintBackRefs{\CurrentBib}

\bibitem [\protect \citeauthoryear {%
Hands%
\ \protect \BOthers {.}}{%
Hands%
\ \protect \BOthers {.}}{%
{\protect \APACyear {2016}}%
}]{%
hands2016new}
\APACinsertmetastar {%
hands2016new}%
\begin{APACrefauthors}%
Hands, A.%
, Lei, F.%
, Ryden, K.%
, Dyer, C.%
, Underwood, C.%
\BCBL {}\ \BBA {} Mertens, C.%
\end{APACrefauthors}%
\unskip\
\newblock
\APACrefYearMonthDay{2016}{}{}.
\newblock
{\BBOQ}\APACrefatitle {New data and modelling for single event effects in the
  stratospheric radiation environment} {New data and modelling for single event
  effects in the stratospheric radiation environment}.{\BBCQ}
\newblock
\APACjournalVolNumPages{IEEE Transactions on Nuclear Science}{64}{1}{587--595}.
\PrintBackRefs{\CurrentBib}

\bibitem [\protect \citeauthoryear {%
Karapetyan%
}{%
Karapetyan%
}{%
{\protect \APACyear {2008}}%
}]{%
karapetyan2008detection}
\APACinsertmetastar {%
karapetyan2008detection}%
\begin{APACrefauthors}%
Karapetyan, G.%
\end{APACrefauthors}%
\unskip\
\newblock
\APACrefYearMonthDay{2008}{}{}.
\newblock
{\BBOQ}\APACrefatitle {Detection of high energy solar protons during ground
  level enhancements} {Detection of high energy solar protons during ground
  level enhancements}.{\BBCQ}
\newblock
\APACjournalVolNumPages{Astroparticle Physics}{30}{5}{234--238}.
\PrintBackRefs{\CurrentBib}

\bibitem [\protect \citeauthoryear {%
Kataoka%
\ \protect \BOthers {.}}{%
Kataoka%
\ \protect \BOthers {.}}{%
{\protect \APACyear {2014}}%
}]{%
kataoka2014radiation}
\APACinsertmetastar {%
kataoka2014radiation}%
\begin{APACrefauthors}%
Kataoka, R.%
, Sato, T.%
, Kubo, Y.%
, Shiota, D.%
, Kuwabara, T.%
, Yashiro, S.%
\BCBL {}\ \BBA {} Yasuda, H.%
\end{APACrefauthors}%
\unskip\
\newblock
\APACrefYearMonthDay{2014}{}{}.
\newblock
{\BBOQ}\APACrefatitle {Radiation dose forecast of WASAVIES during ground-level
  enhancement} {Radiation dose forecast of wasavies during ground-level
  enhancement}.{\BBCQ}
\newblock
\APACjournalVolNumPages{Space Weather}{12}{6}{380--386}.
\PrintBackRefs{\CurrentBib}

\bibitem [\protect \citeauthoryear {%
Lantos%
\ \BBA {} Fuller%
}{%
Lantos%
\ \BBA {} Fuller%
}{%
{\protect \APACyear {2003}}%
}]{%
lantos2003history}
\APACinsertmetastar {%
lantos2003history}%
\begin{APACrefauthors}%
Lantos, P.%
\BCBT {}\ \BBA {} Fuller, N.%
\end{APACrefauthors}%
\unskip\
\newblock
\APACrefYearMonthDay{2003}{}{}.
\newblock
{\BBOQ}\APACrefatitle {History of the solar particle event radiation doses
  on-board aeroplanes using a semi-empirical model and Concorde measurements}
  {History of the solar particle event radiation doses on-board aeroplanes
  using a semi-empirical model and concorde measurements}.{\BBCQ}
\newblock
\APACjournalVolNumPages{Radiation Protection Dosimetry}{104}{3}{199--210}.
\PrintBackRefs{\CurrentBib}

\bibitem [\protect \citeauthoryear {%
Lantos%
\ \BBA {} Fuller%
}{%
Lantos%
\ \BBA {} Fuller%
}{%
{\protect \APACyear {2004}}%
}]{%
lantos2004semi}
\APACinsertmetastar {%
lantos2004semi}%
\begin{APACrefauthors}%
Lantos, P.%
\BCBT {}\ \BBA {} Fuller, N.%
\end{APACrefauthors}%
\unskip\
\newblock
\APACrefYearMonthDay{2004}{}{}.
\newblock
{\BBOQ}\APACrefatitle {Semi-empirical model to calculate potential radiation
  exposure on board airplane during solar particle events} {Semi-empirical
  model to calculate potential radiation exposure on board airplane during
  solar particle events}.{\BBCQ}
\newblock
\APACjournalVolNumPages{IEEE transactions on plasma
  science}{32}{4}{1468--1477}.
\PrintBackRefs{\CurrentBib}

\bibitem [\protect \citeauthoryear {%
Larsen%
, Mishev%
\BCBL {}\ \BBA {} Usoskin%
}{%
Larsen%
\ \protect \BOthers {.}}{%
{\protect \APACyear {2023}}%
}]{%
larsen2023new}
\APACinsertmetastar {%
larsen2023new}%
\begin{APACrefauthors}%
Larsen, N.%
, Mishev, A.%
\BCBL {}\ \BBA {} Usoskin, I.%
\end{APACrefauthors}%
\unskip\
\newblock
\APACrefYearMonthDay{2023}{}{}.
\newblock
{\BBOQ}\APACrefatitle {A New Open-Source Geomagnetosphere Propagation Tool
  (OTSO) and Its Applications} {A new open-source geomagnetosphere propagation
  tool (otso) and its applications}.{\BBCQ}
\newblock
\APACjournalVolNumPages{Journal of Geophysical Research: Space
  Physics}{128}{3}{e2022JA031061}.
\PrintBackRefs{\CurrentBib}

\bibitem [\protect \citeauthoryear {%
Latocha%
, Beck%
\BCBL {}\ \BBA {} Rollet%
}{%
Latocha%
\ \protect \BOthers {.}}{%
{\protect \APACyear {2009}}%
}]{%
latocha2009avidos}
\APACinsertmetastar {%
latocha2009avidos}%
\begin{APACrefauthors}%
Latocha, M.%
, Beck, P.%
\BCBL {}\ \BBA {} Rollet, S.%
\end{APACrefauthors}%
\unskip\
\newblock
\APACrefYearMonthDay{2009}{}{}.
\newblock
{\BBOQ}\APACrefatitle {AVIDOS—A software package for European accredited
  aviation dosimetry} {Avidos—a software package for european accredited
  aviation dosimetry}.{\BBCQ}
\newblock
\APACjournalVolNumPages{Radiation Protection Dosimetry}{136}{4}{286--290}.
\PrintBackRefs{\CurrentBib}

\bibitem [\protect \citeauthoryear {%
Lei%
}{%
Lei%
}{%
{\protect \APACyear {{\protect \bibnodate {}}}}%
}]{%
MAIREUK}
\APACinsertmetastar {%
MAIREUK}%
\begin{APACrefauthors}%
Lei, F.%
\end{APACrefauthors}%
\unskip\
\newblock
\APACrefYearMonthDay{{\protect \bibnodate {}}}{}{}.
\newblock
\APACrefbtitle {{Models for Atmospheric Ionising Radiation Effects - MAIRE}.}
  {{Models for Atmospheric Ionising Radiation Effects - MAIRE}.}
\newblock
\APACrefnote{\url{http://maire.uk/maire/}}
\PrintBackRefs{\CurrentBib}

\bibitem [\protect \citeauthoryear {%
Lei%
, Clucas%
, Dyer%
\BCBL {}\ \BBA {} Truscott%
}{%
Lei%
\ \protect \BOthers {.}}{%
{\protect \APACyear {2004}}%
}]{%
lei2004atmospheric}
\APACinsertmetastar {%
lei2004atmospheric}%
\begin{APACrefauthors}%
Lei, F.%
, Clucas, S.%
, Dyer, C.%
\BCBL {}\ \BBA {} Truscott, P.%
\end{APACrefauthors}%
\unskip\
\newblock
\APACrefYearMonthDay{2004}{}{}.
\newblock
{\BBOQ}\APACrefatitle {An atmospheric radiation model based on response
  matrices generated by detailed Monte Carlo simulations of cosmic ray
  interactions} {An atmospheric radiation model based on response matrices
  generated by detailed monte carlo simulations of cosmic ray
  interactions}.{\BBCQ}
\newblock
\APACjournalVolNumPages{IEEE Transactions on Nuclear
  Science}{51}{6}{3442--3451}.
\PrintBackRefs{\CurrentBib}

\bibitem [\protect \citeauthoryear {%
Lei%
, Hands%
, Clucas%
, Dyer%
\BCBL {}\ \BBA {} Truscott%
}{%
Lei%
\ \protect \BOthers {.}}{%
{\protect \APACyear {2005}}%
}]{%
lei2005improvements}
\APACinsertmetastar {%
lei2005improvements}%
\begin{APACrefauthors}%
Lei, F.%
, Hands, A.%
, Clucas, S.%
, Dyer, C.%
\BCBL {}\ \BBA {} Truscott, P.%
\end{APACrefauthors}%
\unskip\
\newblock
\APACrefYearMonthDay{2005}{}{}.
\newblock
{\BBOQ}\APACrefatitle {Improvements to and validations of the QinetiQ
  Atmospheric Radiation Model (QARM)} {Improvements to and validations of the
  qinetiq atmospheric radiation model (qarm)}.{\BBCQ}
\newblock
\BIn{} \APACrefbtitle {2005 8th European Conference on Radiation and Its
  Effects on Components and Systems} {2005 8th european conference on radiation
  and its effects on components and systems}\ (\BPGS\ D3--1).
\PrintBackRefs{\CurrentBib}

\bibitem [\protect \citeauthoryear {%
Matthiä%
, Berger%
, Mrigakshi%
\BCBL {}\ \BBA {} Reitz%
}{%
Matthiä%
\ \protect \BOthers {.}}{%
{\protect \APACyear {2013}}%
}]{%
matthia2013ready}
\APACinsertmetastar {%
matthia2013ready}%
\begin{APACrefauthors}%
Matthiä, D.%
, Berger, T.%
, Mrigakshi, A\BPBI I.%
\BCBL {}\ \BBA {} Reitz, G.%
\end{APACrefauthors}%
\unskip\
\newblock
\APACrefYearMonthDay{2013}{}{}.
\newblock
{\BBOQ}\APACrefatitle {A ready-to-use galactic cosmic ray model} {A
  ready-to-use galactic cosmic ray model}.{\BBCQ}
\newblock
\APACjournalVolNumPages{Advances in Space Research}{51}{3}{329--338}.
\PrintBackRefs{\CurrentBib}

\bibitem [\protect \citeauthoryear {%
Meier%
\ \protect \BOthers {.}}{%
Meier%
\ \protect \BOthers {.}}{%
{\protect \APACyear {2020}}%
}]{%
meier2020radiation}
\APACinsertmetastar {%
meier2020radiation}%
\begin{APACrefauthors}%
Meier, M\BPBI M.%
, Copeland, K.%
, Kl{\"o}ble, K\BPBI E.%
, Matthi{\"a}, D.%
, Plettenberg, M\BPBI C.%
, Schennetten, K.%
\BDBL {}Hellweg, C\BPBI E.%
\end{APACrefauthors}%
\unskip\
\newblock
\APACrefYearMonthDay{2020}{}{}.
\newblock
{\BBOQ}\APACrefatitle {Radiation in the atmosphere—A hazard to aviation
  safety?} {Radiation in the atmosphere—a hazard to aviation safety?}{\BBCQ}
\newblock
\APACjournalVolNumPages{Atmosphere}{11}{12}{1358}.
\PrintBackRefs{\CurrentBib}

\bibitem [\protect \citeauthoryear {%
C.~Mertens%
\ \protect \BOthers {.}}{%
C.~Mertens%
\ \protect \BOthers {.}}{%
{\protect \APACyear {2023}}%
}]{%
mertens2023nairas}
\APACinsertmetastar {%
mertens2023nairas}%
\begin{APACrefauthors}%
Mertens, C.%
, Gronoff, G.%
, Zheng, Y.%
, Petrenko, M.%
, Buhler, J.%
, Phoenix, D.%
\BDBL {}Minow, J.%
\end{APACrefauthors}%
\unskip\
\newblock
\APACrefYearMonthDay{2023}{}{}.
\newblock
{\BBOQ}\APACrefatitle {NAIRAS Model Run-On-Request Service at CCMC} {Nairas
  model run-on-request service at ccmc}.{\BBCQ}
\newblock
\APACjournalVolNumPages{Space Weather}{21}{5}{e2023SW003473}.
\PrintBackRefs{\CurrentBib}

\bibitem [\protect \citeauthoryear {%
C.~Mertens%
\ \protect \BOthers {.}}{%
C.~Mertens%
\ \protect \BOthers {.}}{%
{\protect \APACyear {2009}}%
}]{%
mertens2009development}
\APACinsertmetastar {%
mertens2009development}%
\begin{APACrefauthors}%
Mertens, C.%
, Tobiska, W\BPBI K.%
, Bouwer, D.%
, Kress, B.%
, Wiltberger, M.%
, Solomon, S.%
\BCBL {}\ \BBA {} Murray, J.%
\end{APACrefauthors}%
\unskip\
\newblock
\APACrefYearMonthDay{2009}{}{}.
\newblock
{\BBOQ}\APACrefatitle {Development of nowcast of atmospheric ionizing radiation
  for aviation safety (nairas) model} {Development of nowcast of atmospheric
  ionizing radiation for aviation safety (nairas) model}.{\BBCQ}
\newblock
\BIn{} \APACrefbtitle {1st AIAA Atmospheric and Space Environments Conference}
  {1st aiaa atmospheric and space environments conference}\ (\BPG~3633).
\PrintBackRefs{\CurrentBib}

\bibitem [\protect \citeauthoryear {%
C\BPBI J.~Mertens%
, Meier%
, Brown%
, Norman%
\BCBL {}\ \BBA {} Xu%
}{%
C\BPBI J.~Mertens%
\ \protect \BOthers {.}}{%
{\protect \APACyear {2013}}%
}]{%
mertens2013nairas}
\APACinsertmetastar {%
mertens2013nairas}%
\begin{APACrefauthors}%
Mertens, C\BPBI J.%
, Meier, M\BPBI M.%
, Brown, S.%
, Norman, R\BPBI B.%
\BCBL {}\ \BBA {} Xu, X.%
\end{APACrefauthors}%
\unskip\
\newblock
\APACrefYearMonthDay{2013}{}{}.
\newblock
\APACrefbtitle {NAIRAS aircraft radiation model development, dose climatology,
  and initial validation.} {Nairas aircraft radiation model development, dose
  climatology, and initial validation.}
\newblock
\APACaddressPublisher{}{Wiley Online Library}.
\PrintBackRefs{\CurrentBib}

\bibitem [\protect \citeauthoryear {%
Mishev%
, Adibpour%
, Usoskin%
\BCBL {}\ \BBA {} Felsberger%
}{%
Mishev%
\ \protect \BOthers {.}}{%
{\protect \APACyear {2015}}%
}]{%
mishev2015computation}
\APACinsertmetastar {%
mishev2015computation}%
\begin{APACrefauthors}%
Mishev, A.%
, Adibpour, F.%
, Usoskin, I.%
\BCBL {}\ \BBA {} Felsberger, E.%
\end{APACrefauthors}%
\unskip\
\newblock
\APACrefYearMonthDay{2015}{}{}.
\newblock
{\BBOQ}\APACrefatitle {Computation of dose rate at flight altitudes during
  ground level enhancements no. 69, 70 and 71} {Computation of dose rate at
  flight altitudes during ground level enhancements no. 69, 70 and 71}.{\BBCQ}
\newblock
\APACjournalVolNumPages{Advances in Space Research}{55}{1}{354--362}.
\PrintBackRefs{\CurrentBib}

\bibitem [\protect \citeauthoryear {%
Mishev%
, Koldobskiy%
, Usoskin%
, Kocharov%
\BCBL {}\ \BBA {} Kovaltsov%
}{%
Mishev%
\ \protect \BOthers {.}}{%
{\protect \APACyear {2021}}%
}]{%
mishev2021application}
\APACinsertmetastar {%
mishev2021application}%
\begin{APACrefauthors}%
Mishev, A.%
, Koldobskiy, S.%
, Usoskin, I.%
, Kocharov, L.%
\BCBL {}\ \BBA {} Kovaltsov, G.%
\end{APACrefauthors}%
\unskip\
\newblock
\APACrefYearMonthDay{2021}{}{}.
\newblock
{\BBOQ}\APACrefatitle {Application of the verified neutron monitor yield
  function for an extended analysis of the GLE\# 71 on 17 May 2012}
  {Application of the verified neutron monitor yield function for an extended
  analysis of the gle\# 71 on 17 may 2012}.{\BBCQ}
\newblock
\APACjournalVolNumPages{Space Weather}{19}{2}{e2020SW002626}.
\PrintBackRefs{\CurrentBib}

\bibitem [\protect \citeauthoryear {%
Mishev%
\ \BBA {} Usoskin%
}{%
Mishev%
\ \BBA {} Usoskin%
}{%
{\protect \APACyear {2016}}%
}]{%
mishev2016analysis}
\APACinsertmetastar {%
mishev2016analysis}%
\begin{APACrefauthors}%
Mishev, A.%
\BCBT {}\ \BBA {} Usoskin, I.%
\end{APACrefauthors}%
\unskip\
\newblock
\APACrefYearMonthDay{2016}{}{}.
\newblock
{\BBOQ}\APACrefatitle {Analysis of the ground-level enhancements on 14 July
  2000 and 13 December 2006 using neutron monitor data} {Analysis of the
  ground-level enhancements on 14 july 2000 and 13 december 2006 using neutron
  monitor data}.{\BBCQ}
\newblock
\APACjournalVolNumPages{Solar Physics}{291}{}{1225--1239}.
\PrintBackRefs{\CurrentBib}

\bibitem [\protect \citeauthoryear {%
Normand%
\ \BBA {} Baker%
}{%
Normand%
\ \BBA {} Baker%
}{%
{\protect \APACyear {1993}}%
}]{%
normand1993altitude}
\APACinsertmetastar {%
normand1993altitude}%
\begin{APACrefauthors}%
Normand, E.%
\BCBT {}\ \BBA {} Baker, T.%
\end{APACrefauthors}%
\unskip\
\newblock
\APACrefYearMonthDay{1993}{}{}.
\newblock
{\BBOQ}\APACrefatitle {Altitude and latitude variations in avionics SEU and
  atmospheric neutron flux} {Altitude and latitude variations in avionics seu
  and atmospheric neutron flux}.{\BBCQ}
\newblock
\APACjournalVolNumPages{IEEE transactions on Nuclear
  Science}{40}{6}{1484--1490}.
\PrintBackRefs{\CurrentBib}

\bibitem [\protect \citeauthoryear {%
Sato%
\ \protect \BOthers {.}}{%
Sato%
\ \protect \BOthers {.}}{%
{\protect \APACyear {2018}}%
}]{%
sato2018real}
\APACinsertmetastar {%
sato2018real}%
\begin{APACrefauthors}%
Sato, T.%
, Kataoka, R.%
, Shiota, D.%
, Kubo, Y.%
, Ishii, M.%
, Yasuda, H.%
\BDBL {}Miyoshi, Y.%
\end{APACrefauthors}%
\unskip\
\newblock
\APACrefYearMonthDay{2018}{}{}.
\newblock
{\BBOQ}\APACrefatitle {Real time and automatic analysis program for WASAVIES:
  Warning system for aviation exposure to solar energetic particles} {Real time
  and automatic analysis program for wasavies: Warning system for aviation
  exposure to solar energetic particles}.{\BBCQ}
\newblock
\APACjournalVolNumPages{Space Weather}{16}{7}{924--936}.
\PrintBackRefs{\CurrentBib}

\bibitem [\protect \citeauthoryear {%
Shea%
\ \BBA {} Smart%
}{%
Shea%
\ \BBA {} Smart%
}{%
{\protect \APACyear {2012}}%
}]{%
shea2012space}
\APACinsertmetastar {%
shea2012space}%
\begin{APACrefauthors}%
Shea, M.%
\BCBT {}\ \BBA {} Smart, D.%
\end{APACrefauthors}%
\unskip\
\newblock
\APACrefYearMonthDay{2012}{}{}.
\newblock
{\BBOQ}\APACrefatitle {Space weather and the ground-level solar proton events
  of the 23rd solar cycle} {Space weather and the ground-level solar proton
  events of the 23rd solar cycle}.{\BBCQ}
\newblock
\APACjournalVolNumPages{Space science reviews}{171}{1}{161--188}.
\PrintBackRefs{\CurrentBib}

\bibitem [\protect \citeauthoryear {%
Taber%
\ \BBA {} Normand%
}{%
Taber%
\ \BBA {} Normand%
}{%
{\protect \APACyear {1993}}%
}]{%
taber1993single}
\APACinsertmetastar {%
taber1993single}%
\begin{APACrefauthors}%
Taber, A.%
\BCBT {}\ \BBA {} Normand, E.%
\end{APACrefauthors}%
\unskip\
\newblock
\APACrefYearMonthDay{1993}{}{}.
\newblock
{\BBOQ}\APACrefatitle {Single event upset in avionics} {Single event upset in
  avionics}.{\BBCQ}
\newblock
\APACjournalVolNumPages{IEEE Transactions on Nuclear Science}{40}{2}{120--126}.
\PrintBackRefs{\CurrentBib}

\bibitem [\protect \citeauthoryear {%
Tobiska%
\ \protect \BOthers {.}}{%
Tobiska%
\ \protect \BOthers {.}}{%
{\protect \APACyear {2015}}%
}]{%
tobiska2015advances}
\APACinsertmetastar {%
tobiska2015advances}%
\begin{APACrefauthors}%
Tobiska, W\BPBI K.%
, Atwell, W.%
, Beck, P.%
, Benton, E.%
, Copeland, K.%
, Dyer, C.%
\BDBL {}others%
\end{APACrefauthors}%
\unskip\
\newblock
\APACrefYearMonthDay{2015}{}{}.
\newblock
{\BBOQ}\APACrefatitle {Advances in atmospheric radiation measurements and
  modeling needed to improve air safety} {Advances in atmospheric radiation
  measurements and modeling needed to improve air safety}.{\BBCQ}
\newblock
\APACjournalVolNumPages{Space Weather}{13}{4}{202--210}.
\PrintBackRefs{\CurrentBib}

\bibitem [\protect \citeauthoryear {%
Tsyganenko%
}{%
Tsyganenko%
}{%
{\protect \APACyear {1989}}%
}]{%
tsyganenko1989magnetospheric}
\APACinsertmetastar {%
tsyganenko1989magnetospheric}%
\begin{APACrefauthors}%
Tsyganenko, N\BPBI A.%
\end{APACrefauthors}%
\unskip\
\newblock
\APACrefYearMonthDay{1989}{}{}.
\newblock
{\BBOQ}\APACrefatitle {A magnetospheric magnetic field model with a warped tail
  current sheet} {A magnetospheric magnetic field model with a warped tail
  current sheet}.{\BBCQ}
\newblock
\APACjournalVolNumPages{Planetary and Space Science}{37}{1}{5--20}.
\PrintBackRefs{\CurrentBib}

\bibitem [\protect \citeauthoryear {%
Usoskin%
, Ibragimov%
, Shea%
\BCBL {}\ \BBA {} Smart%
}{%
Usoskin%
\ \protect \BOthers {.}}{%
{\protect \APACyear {2015}}%
}]{%
usoskin2015database}
\APACinsertmetastar {%
usoskin2015database}%
\begin{APACrefauthors}%
Usoskin, I.%
, Ibragimov, A.%
, Shea, M.%
\BCBL {}\ \BBA {} Smart, D.%
\end{APACrefauthors}%
\unskip\
\newblock
\APACrefYearMonthDay{2015}{}{}.
\newblock
{\BBOQ}\APACrefatitle {{Database of ground level enhancements (GLE) of high
  energy solar proton events}} {{Database of ground level enhancements (GLE) of
  high energy solar proton events}}.{\BBCQ}
\newblock
\APACjournalVolNumPages{Proceedings of Science, Proc. of 34th ICRC Hague,
  Netherlands}{30}{}{}.
\PrintBackRefs{\CurrentBib}

\end{thebibliography}

\end{document}